\pdfoutput=1
\documentclass{easychair}
\usepackage[utf8]{inputenc}
\usepackage[T1]{fontenc}
\usepackage[english]{babel}

\usepackage{appendix}
\usepackage{framed}
\usepackage{lmodern}
\usepackage{amssymb}
\usepackage{amsmath}
\usepackage{todonotes}

\usepackage[vlined,ruled,noend]{algorithm2e}
\usepackage{bussproofs}
\EnableBpAbbreviations
\usepackage{tikz}
\usetikzlibrary{shapes.geometric, patterns}
\usetikzlibrary{calc}
\usetikzlibrary{shapes.misc, positioning}
\usepackage{csquotes}
\usepackage{cite,doi}
\usepackage{booktabs}
\usepackage{thmtools} 
\usepackage{thm-restate}

\usepackage{xcolor}
\usepackage{pgfplots}
\usepackage{caption}

\usepackage{xparse}

\usepackage{stmaryrd}

 \newtheorem{theorem}{Theorem}
 
 \newtheorem{lemma}[theorem]{Lemma}
 \newtheorem{definition}[theorem]{Definition}
 \newtheorem{example}{Example}
 
 \newtheorem*{claim}{Claim}

\usepackage{apptools}
\AtAppendix{\counterwithin{theorem}{section}}
\AtAppendix{\counterwithin{lemma}{section}}
\AtAppendix{\counterwithin{definition}{section}}

\newcommand{\ttodo}[4]{\ifthenelse{\equal{#1}{inline}}{\todo[inline, author=#2, color = 
#3]{#4}}{\todo[color=#3]{#2: #4}}}

\newcommand{\wrt}{w.r.t.\ }
\newcommand{\st}{s.t.\ }
\newcommand{\ie}{i.e.,\xspace}
\newcommand{\eg}{e.g.\ }

\newlength{\myl}
\newcommand{\longsquigarrow}[1]{
    \settowidth{\myl}{$~_{#1}$}
    \raisebox{-0.01cm}{\xymatrix@C=\myl{
            {}\ar@{~>}[r]^{~_{#1}}&{}
        }
    }
}

\newcommand{\MP}{\ensuremath{\mathsf{MP}}\xspace}
\newcommand{\MTP}{\ensuremath{\mathsf{MTP}}\xspace}
\newcommand{\un}{\ensuremath{\mathsf{unary}}\xspace}
\newcommand{\bin}{\ensuremath{\mathsf{binary}}\xspace}

\newcommand{\ct}[1]{\ensuremath{\mathsf{vert}(#1)}\xspace}

\newcommand{\PTime}{\textsc{P}\xspace}
\newcommand{\NP}{\textsc{NP}\xspace}

\newcommand{\ExpTime}{\textsc{ExpTime}\xspace}

\newcommand{\NExpTime}{\textsc{NExpTime}\xspace}
\newcommand{\LogSpace}{\textsc{LogSpace}\xspace}

\newcommand{\poly}{\ensuremath{\mathsf{poly}}\xspace}
\renewcommand{\exp}{\ensuremath{\mathsf{exp}}\xspace}
\newcommand{\reasoner}{deriver\xspace}
\newcommand{\Reasoner}{Deriver}

\newcommand{\Reasoners}{Derivers}

\newcommand{\Imc}{\ensuremath{\mathcal{I}}\xspace}
\newcommand{\Jmc}{\ensuremath{\mathcal{J}}\xspace}

\newcommand{\Lmc}{\ensuremath{\mathcal{L}}\xspace}

\newcommand{\Tmc}{\ensuremath{\mathcal{T}}\xspace} 
\newcommand{\Omc}{\ensuremath{\mathcal{O}}\xspace}

\newcommand{\ALC}{\ensuremath{\mathcal{ALC}}\xspace}
\newcommand{\ALCOI}{\ensuremath{\mathcal{ALCOI}}\xspace}
\newcommand{\EL}{\ensuremath{\mathcal{E}\hspace{-0.1em}\mathcal{L}}\xspace}
\newcommand{\ALCH}{\ensuremath{\mathcal{ALCH}}\xspace}
\newcommand{\ELH}{\ensuremath{\mathcal{E}\hspace{-0.1em}\mathcal{LH}}\xspace}
\newcommand{\ELI}{\ensuremath{\mathcal{ELI}}\xspace}
\renewcommand{\L}{\ensuremath{\mathcal{L}}\xspace}

\newcommand{\nats}{\mathbb{N}}

\newcommand{\tup}[1]{\langle #1 \rangle}

\newcommand{\NC}{\ensuremath{\textsf{N}_\textsf{C}}\xspace}
\newcommand{\NR}{\ensuremath{\textsf{N}_\textsf{R}}\xspace}

\newcommand{\sig}[1]{\ensuremath{\textsf{sig}(#1)}\xspace}

\newcommand{\p}{\ensuremath{\mathcal{P}}\xspace}

\newcommand{\R}{\ensuremath{\mathfrak{D}}\xspace}

\newcommand{\el}{\ensuremath{\ell}\xspace}

\newcommand{\ds}{\ensuremath{\mathcal{D}}\xspace}

\newcommand{\true}{\ensuremath{\mathsf{true}}\xspace}
\newcommand{\false}{\ensuremath{\mathsf{false}}\xspace}

\DeclareDocumentEnvironment{block}{O {} }{\noindent \textbf{#1} \quad }{~\\}

\newcommand{\universe}{\ensuremath{\mathcal{U}}\xspace}

\newcommand{\Lethe}{\textsc{Lethe}\xspace}
\newcommand{\Elk}{\textsc{Elk}\xspace}
\newcommand{\Fame}{\textsc{Fame}\xspace}
\newcommand{\CEL}{\textsc{Cel}\xspace}

\newif\ifhideproofs

\newif\iftechnicalReport

\newif\ifplotDataInTikz

\newif\ifappendix

\plotDataInTikztrue 
\hideproofsfalse
\technicalReporttrue 

\ifhideproofs
  \usepackage{environ}
  \NewEnviron{hide}{}

\fi

\title{Finding Small Proofs for Description Logic Entailments:\\ Theory and Practice\iftechnicalReport\\(Extended Technical Report)\fi}
\titlerunning{Small Proofs for DL Entailments}
\author{Christian Alrabbaa, Franz Baader, Stefan Borgwardt,\\  Patrick Koopmann, and Alisa Kovtunova}
\authorrunning{Alrabbaa, Baader, Borgwardt, Koopmann, Kovtunova}
\institute{Institute of Theoretical Computer Science, TU Dresden, Germany}

\begin{document}

\maketitle


\begin{abstract}
Logic-based approaches to AI have the advantage that their behaviour can in principle be explained by providing their users with proofs
for the derived consequences. However, if such proofs get very large, then it may be hard to understand a consequence even if the individual
derivation steps are easy to comprehend. This motivates our interest in finding small proofs for Description Logic (DL) entailments. Instead
of concentrating on a specific DL and proof calculus for this DL, we introduce a general framework in which proofs are represented as
labeled, directed hypergraphs, where each hyperedge corresponds to a single sound derivation 
step.
On the theoretical side, we investigate the complexity of deciding whether a certain consequence has a proof of size at most $n$
along the following orthogonal dimensions: 
(i)~the underlying proof system is polynomial or exponential;
(ii)~proofs may or may not reuse already derived consequences; and
(iii)~the number $n$ is represented in unary or binary.
We have determined the exact worst-case complexity of this decision problem for all but one of the possible combinations
of these options.
On the practical side, we have developed and implemented an approach for generating proofs 
for expressive DLs based on a non-standard reasoning task called forgetting. We have 
evaluated this approach on a set of realistic ontologies and compared the obtained proofs 
with proofs generated by the DL reasoner ELK, finding that forgetting-based proofs are often better w.r.t.\ different measures 
of proof complexity.
\end{abstract}


\section{Introduction}

Embedded or cyber-physical systems that interact autonomously
with the real world, or with users they are supposed to support,
must continuously make decisions based on sensor data, user input,
knowledge they have acquired during runtime as well as knowledge
provided during design-time. To make the behavior of such systems comprehensible,
they need to be able to explain their decisions to the user or,
after something has gone wrong, to an accident investigator.
While systems that use Machine Learning (ML) to interpret sensor data are
very fast and usually quite accurate, their decisions are notoriously hard to
explain, though huge efforts are currently being made to overcome this problem \cite{XAIpaper}.
In contrast, decisions made by reasoning about symbolically represented knowledge
are in principle easy to explain. In particular, if the knowledge is represented
in (some fragment of) first-order (FO) logic, and a decision is made based on the result
of an FO reasoning process, then one can in principle use a formal proof in an
appropriate calculus to explain an entailment. In practice, however, things are not so rosy also in the
symbolic setting. On the one hand, proofs may be very large, and thus
it may be hard to comprehend why the overall entailment holds even if each single derivation step is
easy to follow. On the other hand, single proof steps may also be hard to understand, 
in particular for users that are not experts in logic.  The problem of explaining why a certain consequence 
follows from a given set of sentences has been considered for full FO automated theorem proving since at least 
four decades. Since the resolution proofs generated by high-performance automated theorem provers
are not appropriate for human consumption, approaches transforming them into
proofs in more human-oriented calculi (such as Gentzen's natural deduction calculus \cite{DBLP:conf/ijcai/Lingenfelder89})
have been developed. To alleviate the tedious task of following a huge number of these steps, 
abstractions of proofs that use definitions, lemmas, and more abstract deduction rules have
been investigated~\cite{DBLP:conf/cade/Huang94}. A more detailed description of the huge body of research in this area
is beyond the scope of this paper.

Here we concentrate on explaining the results of DL reasoning using formal proofs, and in particular on how complex it is to find
small proofs. The first work on explaining DL entailments is probably the PhD thesis of McGuinness \cite{DeMc-96},
where the results obtained by the structural subsumption algorithm of the CLASSIC system \cite{BMPAB91} are translated into
proofs in a formal calculus. The thesis also investigates how to create shorter, better understandable proofs by
pruning away unimportant parts. In \cite{DBLP:conf/ecai/BorgidaFH00}, proofs of subsumptions generated by a tableau-based
system are translated into sequent proofs. The authors then investigate how to make the sequent proofs shorter.
More recent work on explaining DL entailment was often focused on computing so-called
justifications, i.e., minimal subsets of the knowledge base (KB) from which the consequence in question follows 
(see, e.g., \cite{ScCo03,BaPS07,Horr-11}). The basic assumption 
is here that, whereas KBs may be very large and have many consequences, a single consequence often follows from a small 
subset of the KB by an easy derivation. While this is true in certain applications \cite{BaaSun-KRMED-08}, in general
it may be quite hard for a user to see without help why a consequence follows from a given 
justification~\cite{DBLP:conf/semweb/HorridgePS10}. 
On the one hand, this has triggered research into assessing the complexity of a justification, i.e., how hard it is to derive the given
consequence from the justification \cite{DBLP:conf/ekaw/NguyenPPW12,HBPS-KBS13}. On the other hand, it has rekindled
the interest in generating proofs appropriate for human consumption. For example, the explanation plugin
for the ontology editor Prot\'eg\'e described in \cite{KaKS-DL17} cannot only produce justfications, but can also display
proofs, provided that proofs of an appropriate form are returned by the employed reasoner, an assumption that is, e.g.,
satisfied by the reasoner ELK \cite{DBLP:journals/jar/KazakovKS14}. While these proofs are represented in
the formal syntax employed by Prot\'eg\'e, the work reported 
in~\cite{DBLP:conf/dlog/SchillerG13,DBLP:conf/ekaw/NguyenPPW12,DBLP:conf/dlog/SchillerSG17}
uses ontology verbalization techniques to translate proofs into natural language text.

Since most of the previous work on generating proofs for DL entailments emphasizes the importance of small proofs,
we investigate in this paper how hard it is to find such proofs. Instead
of concentrating on a specific DL and proof calculus or reasoner for this DL, 
we introduce a general framework in which proofs are represented as
labeled, directed hypergraphs whose hyperedges correspond to single sound derivation steps.
To be more precise, we assume that a reasoner (called \emph{deriver} in this paper, to distinguish it from an
actual implemented system) generates a so-called derivation structure, which consists of possible proof steps, 
from which actual proofs can be derived. For example, if we consider the consequence-based reasoning approaches for the DLs
$\EL$ and $\ELI$ described in \cite{baader_horrocks_lutz_sattler_2017}, then a derivation structure for a given
KB and consequence consists of the (finitely many) instantiated classification rules. The $\EL$ reasoner ELK
actually returns such a derivation structure, but this structure only contains the rule instances that have actually
been used during the reasoning process.

On the theoretical side, we investigate the complexity of deciding whether a certain consequence has a proof of size at most $n$
along three orthogonal dimensions. First, we distinguish between derivers that produce derivation structures of
polynomial or exponential size. For example, there is a polynomial deriver for $\EL$ (e.g., the one implemented by ELK), 
whereas $\ELI$ has an exponential, but no polynomial deriver. Second, we distinguish between arbitrary proofs
and tree-shaped proofs, which may not reuse already derived consequences, but must re-derive them each time they are needed.
Finally, we distinguish between unary and binary coding of the number $n$, which bounds the size of the proof.
Tables summarizing the complexity results shown in this paper can be found in Section~\ref{sec:complexity}
(Table~\ref{tab:results1} and~\ref{tab:results2}). We see that, for general proofs, the above
decision problem is NP-complete even for polynomial derivers and unary coding of numbers. For exponential
derivers, the complexity depends on the coding of the number $n$: it is NP-complete for unary coding, but NExpTime-complete
for binary coding. Interestingly, for tree-shaped proofs the complexity is considerably lower, which is
due to the fact that we can use a Dijkstra-like greedy algorithm to compute minimal tree-shaped proofs.

On the practical side, we compare the time it requires to extract a smallest proof vs.\ a smallest tree-shaped proof from derivation structures extracted from ELK, and our findings match the theoretical complexity results, \ie finding small tree-shaped proofs is easier.
Moreover, we have developed and implemented an approach for generating proofs for 
expressive DLs based on
\emph{forgetting} \cite{DBLP:conf/dlog/KonevWW09,FOUNDATIONS-UI}. 
We have evaluated this approach on a set of realistic ontologies and compared the proofs 
obtained using the forgetting tools LETHE \cite{LETHE} and FAME \cite{FAME}
with proofs generated by the DL reasoner ELK, finding that forgetting-based proofs are often better w.r.t.\ different measures
of proof complexity. In addition to measuring the size of proofs, we also consider measures obtained by applying
the justification complexity measure of \cite{HBPS-KBS13} to the proof steps.

\iftechnicalReport
This technical report extends the original paper by two appendices. Appendix~\ref{appendix:proofs}~(``Theory'') contains the missing proofs from the main part of this paper, while Appendix~\ref{appendix:generated}~(``Practice'') provides additional examples of automatically generated proofs. 
\fi

\section{Preliminaries}


Most of our theoretical discussion applies to arbitrary \emph{logics}~$\Lmc=(S_\Lmc,\models_\Lmc)$, which consist of a set $S_\Lmc$ of \emph{$\Lmc$-sentences} and a \emph{consequence relation} ${\models_\Lmc}\subseteq P(S_\Lmc)\times S_\Lmc$ between \emph{\Lmc-theories}, \ie subsets of \Lmc-sentences, and single \Lmc-sentences.
We assume that $\models_\Lmc$ has a model-based definition, \ie for some definition of \enquote{model}, $\Tmc\models_\Lmc\alpha$ holds iff every model of all elements in~\Tmc is also a model of~$\alpha$.
We also assume that the \emph{size}~$|\alpha|$ of an \L-sentence~$\alpha$ is defined in some way, \eg by the number of symbols in~$\alpha$.
Since \L is usually fixed, we drop the prefix \enquote{\L-} from now on.
For example, \L could be \emph{first-order logic}.
However, we are mainly interested in proofs for DLs, which can be seen as decidable fragments of first-order logic~\cite{BHLS-17}.
In particular, we use specific DLs to show our hardness results.

The syntax of DLs is based on disjoint, countably infinite sets~\NC and~\NR of \emph{concept names}~$A,B,\dots$ and \emph{role names}~$r,s,\dots$, respectively.
Sentences of the DL~$\EL$, called \emph{general concept inclusions (GCIs)}, are of the form $C\sqsubseteq D$, where $C$ and $D$ are \emph{\EL-concepts}, which are built from concept names by applying the constructors $\top$~(\emph{top}), $C\sqcap D$ (\emph{conjunction}), and $\exists r.C$ (\emph{existential restriction} for a role name~$r$).
The DL \ALC extends \EL by the constructors $\bot$~(\emph{bottom}), $C\sqcup D$ (\emph{disjunction}), $\forall r.C$ (\emph{value restriction}), and $\lnot C$ (\emph{negation}).
Both \EL and \ALC can be extended by a new kind of sentences $r\sqsubseteq s$, called \emph{role inclusions}, where~$r$ and~$s$ are role names, and by additional constructors such as \emph{nominals} or \emph{inverse roles}.
These extensions are denoted by the additional letters $\mathcal{H}$, $\Omc$, and $\Imc$, respectively, appended to the name of the logic, resulting in DLs such as $\ALCH$, $\ALCOI$ or $\ELI$. 
In DLs, finite theories are called \emph{TBoxes} or \emph{ontologies}.
%
We use \sig{X} to denote the \emph{signature} of some ontology or sentence~$X$, \ie the set of concept and role names occurring in~$X$.


The semantics of DLs is based on first-order interpretations; for details, see~\cite{BHLS-17}.
In Figure~\ref{fig:cr}, we depict a simplified version of the inference rules for \EL from~\cite{DBLP:journals/jar/KazakovKS14}.
For example, $\{A\sqsubseteq\exists r.B,\ B\sqsubseteq C,\ \exists r.C\sqsubseteq D\}\models A\sqsubseteq D$ is a valid inference in \EL.
Deciding consequences in \EL is \PTime-complete~\cite{BaBL-IJCAI05}, and in \ELI and \ALC it is \ExpTime-complete~\cite{Schi-IJCAI91,BaBL-OWLED08}.
Given a TBox~\Tmc and sentence~$\alpha$, a \emph{justification} for~$\alpha$ in~\Tmc is a minimal subset $\Jmc\subseteq\Tmc$ such that $\Jmc\models\alpha$.
Already for \EL-ontologies, finding a single justification is possible in polynomial time, but there may be exponentially many justifications, and furthermore finding a justification of size $\le n$ is \NP-complete~\cite{DBLP:conf/ki/BaaderPS07}.
\begin{figure}[tb]
  \input{figures/el-calculus}
  \caption{The inference rules for \EL used in \Elk.}
  \label{fig:cr}
\end{figure}


\section{Proofs, \Reasoners\ and Derivation Structures}

While justifications are a popular tool for pinpointing the reasons for an entailment in an ontology, they do not provide deeper information on the reasoning behind the entailment.
In contrast, consequence-based reasoners such as \CEL~\cite{BaLS-IJCAR06} and \Elk~\cite{DBLP:journals/jar/KazakovKS14} derive new consequences using a fixed set of inference rules, and hence could provide proofs like in Figure~\ref{fig:ex:proof-tree}, which uses the inference rules from Figure~\ref{fig:cr}.
\begin{figure}[tb]
\centering
\begin{minipage}[b]{0.4\linewidth}
\centering
\AxiomC{$A \sqsubseteq B$}
\AXC{}
\UIC{$B\sqsubseteq B$}
\AxiomC{$B \sqsubseteq \exists r.A$}
\BinaryInfC{$B \sqsubseteq B \sqcap \exists r.A$}
\BinaryInfC{$A \sqsubseteq B \sqcap \exists r.A$}
\DP
\caption{A tree-shaped proof in \EL}\label{fig:ex:proof-tree}
\end{minipage}
\end{figure}
%
Of course, the precise structure of a proof depends on the inference rules that are allowed to draw conclusions. Moreover, provers usually output a consequence only once, even if it is used in several inference steps. To be able to study abstract properties of such proofs, we view proofs as directed hypergraphs, in which hyperedges represent inference steps that connect a set of sentences (the premises) to another sentence (the conclusion). 
In the following, we fix a logic~\L.

%
\begin{definition}[Hypergraph]
A \emph{(directed, labelled) hypergraph}~\cite{DBLP:journals/cor/NielsenAP05} is a triple $H=(V,E,\el)$, where 
\begin{itemize}
\item $V$ is a finite set of \emph{vertices}, 
\item $E$ is a set of \emph{hyperedges}~$(S,d)$, where $S\subseteq V$ and $d\in V$, and
\item $\el\colon V\to S_\L$ is a \emph{labelling function} that assigns sentences to vertices.
\end{itemize}
\end{definition}
The \emph{size} of~$H$, denoted~$|H|$, is measured by the size of the labels of its hyperedges: $$|H|:=\sum_{(S,d)\in E}\Big(|\el(d)|+\sum_{v\in S}|\el(v)|\Big).$$
%
A vertex $v\in V$ is called a \emph{leaf} if it has no incoming hyperedges, \ie there is no $(S,v)\in E$, and $v$ is a \emph{sink} if it has no outgoing hyperedges, \ie there is no $(S,d)\in E$ such that $v\in S$.

\begin{definition}[Derivation Structure]\label{def:derivation-structure}
  A \emph{derivation structure} $\ds = (V, E, \el)$ over a finite theory~\Tmc is a hypergraph, where
  \begin{itemize}
  \item \ds is \emph{grounded}, \ie 
  every leaf $v$ in~\ds is labeled by $\el(v)\in\Tmc$; and 
%
  \item \ds is \emph{sound}, \ie for every $(S,d)\in E$, the entailment $\{\el(s)\mid s\in S\}\models\el(d)$ holds.
\end{itemize}
A derivation structure $\ds'= (V', E', \el')$ is called a \emph{substructure} of $\ds= (V, E, \el)$ if $V'\subseteq V$, $E'\subseteq E$ and $\el'=\el|_{V'}$. In this case, we also say that \ds \emph{contains} $\ds'$.
\end{definition}
In such a structure, the hyperedges are called \emph{inference steps}.
There can be inference steps of the form $(\emptyset,v)$ if $v$ is labeled by a tautology, for example in Figure~\ref{fig:ex:proof-graph}, where the leafs are marked with a thick border and labeled by the \EL sentences from $\Tmc=\{A\sqsubseteq B, B\sqsubseteq \exists r.A\}$ and the hyperedges represent valid \EL entailments, in particular $\emptyset\models\{B\sqsubseteq B\}$.
This hypergraph correponds to the tree-shaped proof in Figure~\ref{fig:ex:proof-tree}.
On the other hand, Figure~\ref{fig:ex:graph-cycle} depicts a derivation structure that does not correspond to a proof since it contains cyclic inferences.
%
%
%
\begin{figure}[tb]
\centering
\begin{minipage}[b]{0.52\linewidth}
\centering
\begin{tikzpicture}[scale=0.8,
block/.style={
draw,
rounded rectangle, 
minimum width={width("$A \sqsubseteq \exists r.(\exists r.A \sqcap B)$")-5pt}},
he/.style={rounded corners=10pt,->}]
\path (7,3) node[block, very thick] (x1) {$B \sqsubseteq \exists r.A$}
		(3,3) node[block] (x2) {$B \sqsubseteq B$}
		(5,2) node[block] (x3) {$B \sqsubseteq B \sqcap \exists r.A$}
		(1,2) node[block,very thick] (y) {$A \sqsubseteq B$}
		(3,1) node[block] (y3) {$A \sqsubseteq B \sqcap \exists r.A$};
\draw[he] (x1) -- ($(x1)!0.5!(x2)$) -- (x3);
\draw[he] (x2) -- ($(x1)!0.5!(x2)$) -- (x3);
\draw[he] (y) -- ($(y)!0.5!(x3)$) -- (y3);	
\draw[he] (x3) -- ($(y)!0.5!(x3)$) -- (y3);	
\draw[he] ($(x2)+(0,0.7)$) -- (x2);
\end{tikzpicture}
\caption{A derivation structure (which is a proof)}\label{fig:ex:proof-graph}
\end{minipage}
\hspace{0.2cm}
\begin{minipage}[b]{0.45\linewidth}
\centering
\begin{tikzpicture}[scale=0.8,
block/.style={
draw,
rounded rectangle, 
minimum width={width("$A \sqsubseteq \exists r.(\exists r.A \sqcap B)$")-5pt}},
he/.style={rounded corners=10pt,->}]
\path (5,2) node[block] (x3) {$A \sqsubseteq \exists r.A$}
		(1,2) node[block] (y) {$A \sqsubseteq B$}
		(3,1) node[block] (y3) {$A \sqsubseteq B \sqcap \exists r.A$};
		%
\draw[he] (y3) -- (y3-|y) -- (y);
\draw[he] (y3) -- (y3-|x3) -- (x3);
\draw[he] (y) -- ($(y)!0.5!(x3)$) -- (y3);
\draw[he] (x3) -- ($(y)!0.5!(x3)$) -- (y3);
\end{tikzpicture}
\caption{A cyclic derivation structure}\label{fig:ex:graph-cycle}
\end{minipage}
\end{figure}

\begin{definition}[Cycle, Tree]
Given a hypergraph $H=(V,E,\el)$ and $s,t\in V$, a \emph{path} $P$ of length $q$ in~$H$ from~$s$ to~$t$ is a sequence of vertices and
hyperedges
\[ P=(d_0,(S_1,d_1),d_1,(S_2,d_2),\dots, d_{q-1},(S_q,d_q),d_q), \]
where $d_0=s$, $d_q=t$, and $d_{j-1}\in S_j$ for all $j$, $1\le j\le q$. If there is such a path in~$H$, we say that $t$ is \emph{reachable} from~$s$ in~$H$. 
If $t = s$, then $P$ is called a \emph{cycle}.
A hypergraph is \emph{acyclic} if it does not contain a cycle.

A \emph{tree} $H=(V,E,\el)$ with \emph{root} $t\in V$ is a hypergraph in which $t$ is a sink and is reachable from every vertex $v\in V\setminus\{t\}$ by exactly one path.
\end{definition}
In particular, the root is the only sink in a tree, and all trees are acyclic.
%
%
We can now define proofs as special derivation structures that derive a goal sentence. 
\begin{definition}[Proof]\label{def:proof}
    %
%
%
%
%
  Given a sentence~$\eta$ and a finite theory~\Tmc, a \emph{proof for $\Tmc\models\eta$} is a derivation structure $\p = (V, E,\el)$ over~\Tmc that
\begin{itemize}
\item\label{item1:def-proof-non-redundancy}    
contains exactly one sink~$v_\eta\in V$, which is labelled by~$\eta$, and
\item
  is acyclic.
\end{itemize}
A \emph{tree proof} is a proof that is a tree.
\end{definition}
%
For example, the derivation structure in Figure~\ref{fig:ex:proof-graph} is a proof according to this definition, but the one in Figure~\ref{fig:ex:graph-cycle} is not.
%
%
%
%
Figures~\ref{fig:ex:proof-tree-2} and~\ref{fig:ex:proof-graph-2} depict two more ways to prove $A\sqsubseteq B\sqcap\exists r.A$ from $\Tmc=\{ A \sqsubseteq B,\ B \sqsubseteq \exists r.A \}$ (following the inference rules in Figure~\ref{fig:cr}). The first is presented in classical notation as a tree and the second one as a hypergraph without sentence label repetition.
Both of them can be seen as proofs in the sense of Definition~\ref{def:proof} and they use the same inference steps, but have different numbers of vertices.
%
\begin{figure}[tb]
\begin{minipage}[b]{0.45\linewidth}
\centering
%
\AxiomC{$A \sqsubseteq B$}
\AxiomC{$A \sqsubseteq B$}
\AxiomC{$B \sqsubseteq \exists r.A$}
\BinaryInfC{$A \sqsubseteq \exists r.A$}
\BinaryInfC{$A \sqsubseteq B \sqcap \exists r.A$}
\DP
\caption{A tree-shaped proof}\label{fig:ex:proof-tree-2}
%
\end{minipage}
\hspace{0.5cm}
\begin{minipage}[b]{0.45\linewidth}
\centering
\begin{tikzpicture}[scale=0.8,
block/.style={
draw,
rounded rectangle,
minimum width={width("$C \sqsubseteq \exists r.(\exists r.A \sqcap B)$")-5pt}},
he/.style={rounded corners=10pt,->}]
\path (7,3) node[block, very thick] (x1) {$B \sqsubseteq \exists r.A$}
		(3,3) node[block, very thick] (x2) {$A \sqsubseteq B$}
		(5,2) node[block] (x3) {$A \sqsubseteq \exists r.A$}
		(3,1) node[block] (z) {$A \sqsubseteq B \sqcap \exists r.A$};
		%
\draw[he] (x1) -- ($(x1)!0.5!(x2)$) -- (x3);
\draw[he] (x2) -- ($(x1)!0.5!(x2)$) -- (x3);
\draw[he] (x3.west) -- (x3-|z) -- (z);
\draw[he] (x2) -- (z);
\end{tikzpicture}
\caption{A proof with a reused vertex}\label{fig:ex:proof-graph-2}
\end{minipage}
\end{figure}

To compare these different kinds of proofs, we recall the notion of homomorphism.

\begin{definition}[Homomorphism]\label{def:homomorphism}
Let $H=(V,E,\el)$ and $H'=(V',E',\el')$ be two hypergraphs.  A \emph{homomorphism} from $H$ to $H'$, denoted $h\colon H\rightarrow H'$, is a mapping $h\colon V\to V'$ such that for all $(S,d)\in E$, one has $h(S,d):=(\{h(v)\mid v\in S\},h(d))\in E'$ and, for all $v\in V$, it holds that $\el'(h(v))=\el(v)$.
The homomorphism~$h$ is an \emph{isomorphism} if it is a bijection and, for all $S\subseteq V$ and $d\in V$, we have $(S,d)\in E$ iff $h(S,d)\in E'$.
\end{definition}


Recall that our goal is to find, among all possible proofs of a sentence using a specific set of inference rules, a \enquote{minimal} one.
Before we define what exactly we mean by \enquote{minimal,} we first need to specify what \enquote{all possible proofs} means.
For this, we assume that we are given a \emph{\reasoner} (\eg a first-order theorem prover or a DL reasoner) that, given a theory~\Tmc and a goal sentence~$\eta$, produces a derivation structure~\ds.
This structure describes all instances of inference rules that are relevant for constructing proofs for~$\eta$, without us having to know the details of these rules.
The quest for a minimal proof thus becomes a search for a minimal proof expressible by the inference steps in~\ds.
Since systems of inference rules may not be complete for proving arbitrary sentences of a given logic~\L, we restrict $\eta$ in the following definition to a subset $C_\L\subseteq S_\L$ of allowed consequences.
For example, in DLs a common reasoning task is \emph{classification}, which amounts to checking entailment of atomic concept subsumptions of the form $A\sqsubseteq B$, where $A$ and $B$ are concept names.
%
\begin{definition}[\Reasoner{}]\label{def:poly-exp-ds}
A \emph{\reasoner}~\R is given by a set $C_\L\subseteq S_\L$ and a function
that assigns derivation structures to pairs $(\Tmc,\eta)$ of finite theories~$\Tmc\subseteq S_\L$ and sentences~$\eta\in C_\L$, such that
$\Tmc\models\eta$ iff $\R(\Tmc,\eta)$ contains a proof for~$\Tmc\models\eta$.
A proof~\p for $\Tmc\models\eta$ is called \emph{admissible \wrt $\R(\Tmc,\eta)$} if there is a homomorphism $h\colon\p\to\R(\Tmc,\eta)$.


We call $\R$ a \emph{polynomial \reasoner} if there exists a polynomial~$p(x)$ such that the size of $\R(\Tmc,\eta)$ is bounded by $p(|\Tmc|+|\eta|)$. \emph{Exponential \reasoner{}s} are defined similarly by the restriction $|\R(\Tmc,\eta)|\le 2^{p(|\Tmc|+|\eta|)}$.

\end{definition}
%


\Elk~\cite{DBLP:journals/jar/KazakovKS14} can be seen as a polynomial \reasoner{} for \EL in this sense, because it allows to instantiate the inference rules in Figure~\ref{fig:cr} with only a polynomial number of \EL-concepts, namely the subconcepts appearing in~\Tmc or~$\eta$. 
The derivation structure $\text{\Elk}(\Tmc,\eta)$ thus contains all allowed instances of these rules, where each sentence is represented by a unique vertex.
Since the number of premises in each rule is bounded by~$2$, the overall size of this structure is polynomial.
\Elk is complete only for some goal sentences of the form $C\sqsubseteq D$, where $D$ is a concept name and $C$ is a subconcept from~\Tmc.%
\footnote{To prove other kinds of entailments $\Tmc\models\eta$, one first has to \emph{normalize} \Tmc and $\eta$~\cite{BaBL-IJCAI05}.}
%
%
%
An example of an exponential \reasoner{} is given by the inference rules 
for~\ALC~\cite{DBLP:conf/ijcai/SimancikKH11}, which are complete for entailments of the form 
$\Tmc\models M\sqsubseteq\bot$, where \Tmc is normalized 
(see~\cite{DBLP:conf/ijcai/SimancikKH11} for details) and $M$ is a conjunction of literals 
(concept names or negated concept names).
For each such \Tmc and $M$, it is guaranteed that there are only exponentially many valid 
inference steps.
Other examples include
a triple-exponential inference system for the very expressive DL $\mathcal{ALCHOIQ}$~\cite{DBLP:conf/ijcai/CucalaGH18} and an exponential \reasoner{} for \ELI~\cite{baader_horrocks_lutz_sattler_2017}. For a survey on consequence-based reasoning in DLs, see \cite{DBLP:conf/birthday/CucalaGH19}.

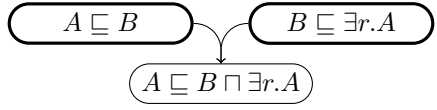
\begin{figure}[tb]
\centering
\begin{tikzpicture}[scale=0.8,
block/.style={
draw,
rounded rectangle,
minimum width={width("$C \sqsubseteq \exists r.(\exists r.A \sqcap B)$")-5pt}},
he/.style={rounded corners=10pt,->}]
\path (7,3) node[block, very thick] (x1) {$B \sqsubseteq \exists r.A$}
		(3,3) node[block, very thick] (x2) {$A \sqsubseteq B$}
		(5,2) node[block] (x3) {$A \sqsubseteq B \sqcap \exists r.A$};
\draw[he] (x1) -- ($(x1)!0.5!(x2)$) -- (x3);
\draw[he] (x2) -- ($(x1)!0.5!(x2)$) -- (x3);
\end{tikzpicture}
\caption{A justification proof}\label{fig:justification}
\end{figure}

Note that the generality of our definitions also admits proofs as in Figure~\ref{fig:justification}, where the consequence is derived directly from its justification in one inference step.
Indeed, such a short proof can sometimes be the most effective explanation.
In this paper, we leave the choice of which granularity of inference steps is most appropriate to the chosen \reasoner.


\section{The Complexity of Finding Minimal Proofs}
\label{sec:complexity}

We focus on two kinds of proof representations (cf.\ Figures~\ref{fig:ex:proof-tree-2} and~\ref{fig:ex:proof-graph-2}).
In Section~\ref{sec:trees}, we consider tree-shaped proofs in the classical form that is naturally suited for a two-dimensional layout.
However, by allowing an arbitrary hypergraphs as in Section~\ref{sec:hypergraphs}, one may obtain more concise proofs.
This is closer to the usual output of an automated inference system, which reports each derived consequence only once.
These two approaches have important differences which result in different complexity bounds for the problem of finding a minimal proof.
Due to lack of space, we include only proof sketches and defer the complete proofs to the 
\iftechnicalReport
    appendix.
\fi


\subsection{Minimal Proofs}\label{sec:hypergraphs}

We formally define our main problem: finding minimal admissible proofs \wrt $\R(\Tmc,\eta)$. 
We consider a proof minimal if it has a minimal number of vertices. Given a proof $\p=(V,E,\el)$, 
we define $\ct{\p}:=|V|$ for any proof $\p=(V,E,\el)$. 

\begin{definition}[Minimal Proof]
\label{def:mp}
  Let \R be a \reasoner{}.
  Given a theory~\Tmc and a sentence~$\eta\in C_\L$, an admissible proof \p  \wrt 
  $\R(\Tmc,\eta)$ is called \emph{minimal} if $\ct{\p}$ is minimal among all such proofs.
  The associated decision problem, denoted $\MP(\R)$, is to decide, given \Tmc and~$\eta$ as above and a natural number $n\in\nats$, whether there is an admissible proof~\p \wrt $\R(\Tmc,\eta)$ with $\ct{\p}\le n$.
  
  If \R is a polynomial \reasoner{}, we add the superscript~$\poly$ to $\MP$, in case of an exponential \reasoner{} we add~$\exp$. If $n$ is encoded in unary representation, we add a subscript~$\un$, and for binary encoding~$\bin$, \eg $\MP^\poly_\un(\R)$.
\end{definition}

Note that an admissible proof~\p has no less vertices than its homomorphic image in $\R(\Tmc,\eta)$.
We can show that if \p is minimal, then this image must be isomorphic to~\p, and therefore is also a proof.
Thus, to decide $\MP(\R)$ we can focus on proofs which are substructures of $\R(\Tmc,\eta)$.

\begin{restatable}{lemma}{hyperproofInside}\label{lem:adm-hyperproof-inside}
There exists a substructure in $\R(\Tmc,\eta)$ which is a minimal proof.
\end{restatable}

Our complexity results for $\MP(\R)$ are summarized in Table~\ref{tab:results1}.
\begin{table}[tb]
\begin{minipage}{0.52\textwidth}
\centering
\caption{The complexity of $\MP(\R)$.}
\label{tab:results1}
\begin{tabular}{c@{\quad}c@{\quad}c}
\toprule
& \poly & \exp \\
\midrule
\un & $\NP$ \scriptsize{[Th.\,\ref{th:poly-graph-NPhard}]} & $\NP$  \scriptsize{[Th.\,\ref{th:exp-unary-NP-mem}]} \\
\bin & $\NP$ \scriptsize{[Th.\,\ref{th:poly-NP-exp-NExp-mem}]} & $\NExpTime$ \scriptsize{[Th.\,\ref{th:poly-NP-exp-NExp-mem}, Th.\,\ref{th:exp-graph-NExpHard}]} \\
\bottomrule
\end{tabular}
\end{minipage}
\begin{minipage}{0.46\textwidth}
\centering
\caption{The complexity of $\MTP(\R)$.}
\label{tab:results2}
\begin{tabular}{c@{\quad}c@{\quad}c}
\toprule
& \poly & \exp \\
\midrule
\un & $\PTime$ \scriptsize{[Th.\,\ref{th:MTP-poly-P-hard}]} & $\NP$ \scriptsize{[Th.\,\ref{th:MTP-exp-NP-hard}, Th.\,\ref{th:MTP-exp-unary-NP-mem}]} \\
\bin & $\PTime$ \scriptsize{[Th.\,\ref{th:MTP-poly-P-mem}]} & $\le \ExpTime$ \scriptsize{[Th.\,\ref{th:MTP-exp-ExpTime-mem}]} 
 \\
\bottomrule
\end{tabular}
\end{minipage}
\end{table}

To show the upper bounds for the second row of the table, we can simply guess a substructure~\p of $\R(\Tmc,\eta)$ and check the conditions of Definitions~\ref{def:derivation-structure} and~\ref{def:proof} and that $\ct{\p}\le n$.

\begin{restatable}{theorem}{thPolyNPExpNExpMem}\label{th:poly-NP-exp-NExp-mem}
  $\MP^\poly_\bin(\R)$ is in \NP and $\MP^\exp_\bin(\R)$ is in \NExpTime.
\end{restatable}

For the case of~$\MP^\exp_\un(\R)$, we can show an \NP-upper bound if we make some additional (reasonable) assumptions. In practice, in order to search for a proof of polynomially bounded size, it does not make sense to construct a complete derivation structure of exponential size first. Instead, a reasoning system would request relevant parts of the derivation structure on demand, for instance guided by a calculus. As a simple abstraction of this, for our complexity upper bounds, 
%
%
we assume that we can access derivation structures by means of \emph{oracle functions} that can instantly check whether a given pair $(S,d)$ is a hyperedge in $\R(\Tmc,\eta)$, and whether a given vertex has a certain label.%
\footnote{This is similar to how a random-access Turing machine can access its input tape by specifying cell addresses.}
%

\begin{restatable}{theorem}{thExpUnaryNPMem}
\label{th:exp-unary-NP-mem}
  $\MP^\exp_\un(\R)$ is in \NP.
\end{restatable}

The \NP-hardness of $\MP^\poly_\un(\R)$ can be shown by a reduction of the hitting set problem~\cite{10.5555/574848} using \Elk with the inference rules in Figure~\ref{fig:cr} as a polynomial \reasoner{}. The hardness argument is inspired by a similar result for justifications in~\cite{DBLP:conf/ki/BaaderPS07}.

\begin{restatable}{theorem}{thPolyGraphNPhard}\label{th:poly-graph-NPhard}
  There is a polynomial \reasoner~\R such that $\MP^\poly_\un(\R)$ is \NP-hard.
\end{restatable}

It remains to show one lower bound for Table~\ref{tab:results1}. 
For $\MP^\exp_\bin(\R)$, we can use the \reasoner given by the calculus for \ELI from~\cite{baader_horrocks_lutz_sattler_2017}, and perform a reduction from a $\NExpTime$-hard tiling problem. While the reduction is relatively complex, the main idea can be briefly sketched as follows: we use the sentences in the derivation structure to refer to possible combinations of coordinates in the tiling with associated tile types, where two roles are used to refer to possible neighbours. The entailment to be proven requires to infer at least the sentences in a possible tiling. The bound on the size of the proof makes sure that the same coordinates cannot be used twice, thus identifying coordinates with tile types. A special challenge arises from the fact that even auxiliary sentences and inferences contribute to the proof size, which should not depend on specifics of the tiling solutions such as whether all tiling conditions are used or not.
%
\begin{restatable}{theorem}{thExpGraphNExpHard}\label{th:exp-graph-NExp-hard}
\label{th:exp-graph-NExpHard}
There is an exponential \reasoner~\R such that $\MP^\exp_\bin(\R)$ is \NExpTime-hard.
\end{restatable}


\subsection{Minimal Tree Proofs}
\label{sec:trees}

We now introduce a variant of $\MP(\R)$ that considers tree-shaped proofs. 

\begin{definition}[Minimal Tree Proof]
\label{def:mtp}
  Let \R be a \reasoner{}. Given a theory~\Tmc and a sentence~$\eta\in C_\L$ such that $\Tmc\models\eta$, an admissible tree proof \wrt $\R(\Tmc,\eta)$ is called \emph{minimal} if \ct{\p} is minimal among all such tree proofs.
  We use $\MTP(\R)$ analogously to $\MP(\R)$.
\end{definition}

Table~\ref{tab:results2} summarizes our results regarding $\MTP(\R)$. Surprisingly, for polynomial \reasoner{}s the complexity drops from~\NP to~\PTime when allowing only tree proofs.
The reason is that we can use a greedy, Dijkstra-like algorithm \cite{DBLP:journals/dam/GalloLP93,DBLP:journals/cor/NielsenAP05} to find a minimal tree proof.
The original algorithm is designed to find shortest hyperpaths, which are defined as follows~\cite{DBLP:journals/cor/NielsenAP05}.

\begin{definition}[Hyperpath]\label{def:hyperpath}
Given two vertices~$s$ and~$t$, a \emph{hyperpath} $\Pi$ is an acyclic 
hypergraph
$\Pi=(V,E,\el)$ such that:
\begin{itemize}
\item $s,t\in V$, 
\item for any $v\in V\setminus \{s\}$, $v$ is reachable from~$s$ in~$\Pi$, and
\item there exists no proper substructure of $\Pi$ with the above properties.
\end{itemize}
\end{definition}
The second condition implies that for each $v\in V\setminus \{s\}$ there is an incoming hyperedge $(S,v)$ in~$\Pi$, while the third implies that hyperedge $(S,v)$ is unique, \ie there is no other incoming hyperedge for $v$.

%
Given a derivation structure $\ds=(V,E,\el)$ over a theory~\Tmc, we define $\ds^*=(V\cup \{s\},E^*,\el^*)$ by extending $\ds$ with an auxiliary vertex~$s$, whose label is arbitrary, and new edges connecting $s$ to every vertex labelled by a tautology or an element of~\Tmc:
$$
E^*=E\cup \{(\{s\},v) \mid v\in V,\,\el(v)\in \Tmc\} \cup \{(\{s\},v) \mid (\emptyset,v)\in E\}.
$$

Intuitively, proofs for $\Tmc\models\eta$ in~\ds correspond to hyperpaths in~$\ds^*$ that 
connect~$s$ to a target vertex~$t$ labeled by~$\eta$, and tree proofs correspond to the 
unraveling of such hyperpaths.%
\footnote{Similar to graph theory, by the unraveling (unfolding) of a hypergraph we mean a (subset-)minimal tree hypergraph obtained from the original hypergraph by copying vertices, such that for each path in the original hypergraph there is a corresponding path, up to copies, in the unraveled hypergraph.
}
However, we also need to be able to measure the size of a tree proof based on the corresponding hyperpath.
The algorithm from~\cite{DBLP:journals/dam/GalloLP93} finds a hyperpath~$\p=(V,E,\el)$ in $\ds^*$ from~$s$ to~$t$ that is minimal \wrt a so-called \emph{additive weighting function}.%
\footnote{In the case of non-additive functions, \cite{italiano1989online} have proved that the problem of finding minimal hyperpaths is \NP-hard, which is analogous to the result in Theorem~\ref{th:poly-graph-NPhard}.}
We are here interested in the following such function:
\begin{align*}
W_\p(s)&:=0, &&\text{for the source vertex }s, \\
W_\p(d)&:=1+\sum_{v\in S} W_\p(v), &&\text{for every }(S,d)\in E, \\
W_\p(\p)&:=W_\p(t), &&\text{for the target vertex }t.
\end{align*}
This is well-defined since a hyperpath can contain no cycles and exactly one incoming edge for each vertex except~$s$. Intuitively, additive weighting functions (corresponding to minimal tree proofs) do not need to keep track of whether a particular vertex/sentence has already been visited before, but simply add a value---in our case~$1$---whenever the vertex is visited. In other words, admissible tree proofs can use multiple copies of vertices from the derivation structure $\R(\Tmc,\eta)$. In contrast, for minimal proofs in general (corresponding to special non-additive weighting functions) one has to keep track of which vertices/sentences have already been visited, in order to avoid counting them twice.

\begin{restatable}{lemma}{lemTreeProofsHyperpaths}\label{lem:tp-hyp}
  There is a tree proof~\p for~$\Tmc\models\eta$ that is admissible \wrt $\ds=\R(\Tmc,\eta)$ with $\ct{\p}\le m$ iff $\ds^*$ contains a hyperpath~$\p'$ from~$s$ to a vertex~$t$ labeled with~$\eta$ with $W_{\p'}(\p')\le m$.
\end{restatable}

Hence, the polynomial algorithm \emph{SBT-Dijkstra} from~\cite{DBLP:journals/dam/GalloLP93} can be used to find minimal tree proofs in polynomial time in the size of the derivation structure $\R(\Tmc,\eta)$. 

\begin{restatable}{theorem}{thMTPPolyPMem}\label{th:MTP-poly-P-mem}\label{th:MTP-exp-ExpTime-mem}
  $\MTP^\poly_\bin(\R)$ is in \PTime and $\MTP^\exp_\bin(\R)$ is in \ExpTime.
\end{restatable}

For the matching lower bound for $\MTP^\poly_\un(\R)$, we reduce the \PTime-hard entailment problem in \EL without existential restrictions~\cite{DBLP:journals/jlp/DowlingG84} to $\MTP(\R)$, where \R is \Elk.
The challenge here is to show that minimal tree proofs admissible \wrt this derivation structure are bounded polynomially (which trivially holds in the case of $\MP(\Elk)$).
For that we propose a strategy for applying the inference from Figure~\ref{fig:cr} (omitting existential restrictions) which demonstrates that there is no superpolynomially large minimal tree proof admissible \wrt the \reasoner{}.
%
\begin{restatable}{theorem}{thMTPPolyPHard}\label{th:MTP-poly-P-hard}
  There exists a polynomial \reasoner~\R such that
  $\MTP^\poly_\un(\R)$ is \PTime-hard under \LogSpace-reductions.
\end{restatable}

Regarding the lower bound for $\MTP^\exp_\un(\R)$, we again use the exponential deriver for \ELI from~\cite{baader_horrocks_lutz_sattler_2017}, and this time use it for a reduction of SAT. Different to the lower bounds in the last subsection, we cannot use the number of vertices to restrict the usage of sentences throughout the whole proof: the special property of tree proofs is that sub-trees cannot be shared among different parts of the proof. The solution is to build the TBox in such a way that minimal proofs are mostly linear. For SAT, we need to simulate the ``guessing'' of a valuation for the SAT formula, which is then checked against the different clauses of the input formula. The intuitive idea is to collect the truth values in a single concept using a linear sequence of rule applications, and use another linear sequence to check this concept against each clause one after the other. The number of vertices is used to restrict the non-deterministic choices in the proof, and allows us to enforce that, for each variable, only one valuation is picked.


\begin{restatable}{theorem}{thMTPExpNPHard}\label{th:MTP-exp-NP-hard}
There is an exponential \reasoner~\R such that $\MTP^\exp_\un(\R)$ is \NP-hard.
\end{restatable}

The remaining \NP upper bound for Table~\ref{tab:results2} can be shown analogously to Theorem~\ref{th:exp-unary-NP-mem}.

\begin{restatable}{theorem}{thMTPExpUnaryNPMem}\label{th:MTP-exp-unary-NP-mem}
  $\MTP^\exp_\un(\R)$ is in \NP.
\end{restatable}
%


\section{Finding Proofs in Practice}
\label{sec:evaluation}

\tikzset{marks/.style={only marks, mark size = 0.4pt, solid, fill=red, line width=0pt, opacity=1, draw opacity=0}}

\newcommand{\plotData}[6]{
    \begin{tikzpicture}
        \tiny
        \begin{axis}[height=0.35\textwidth, 
               width=0.35\textwidth,
               title={#1},
               xlabel={#3},
               ylabel={#4},
               every axis title/.append style={
                 at={(0.5,.95)}
               },
               every axis y label/.style={
                    at={(ticklabel cs:0.5)},rotate=90, anchor=near ticklabel,
               },
               every axis x label/.style={
                    at={(ticklabel cs:0.5)}, anchor=near ticklabel,
               },
               xmin=0,
               xmax=#5,
               ymin=0,
               ymax=#5,
               domain = 0:#5,
               grid=major,
               xtick distance=#6,
               ytick distance=#6,
              ]
            \addplot[marks] table [y index = 1, x index = 0, white space        chars={{,},\ }]  {#2}; 
            \addplot[draw=gray!50!white] {x};
        \end{axis}
    \end{tikzpicture}
}

\newcommand{\plotDataLog}[6]{
    \begin{tikzpicture}
        \tiny
        \begin{axis}[height=0.35\textwidth, 
               width=0.35\textwidth,
               title={#1},
               xlabel={#3},
               ylabel={#4},
               ymode=log,
               xmode=log,
               every axis title/.append style={
                 at={(0.5,.95)}
               },
               every axis y label/.style={
                    at={(ticklabel cs:0.5)},rotate=90, anchor=near ticklabel,
               },
               every axis x label/.style={
                    at={(ticklabel cs:0.5)}, anchor=near ticklabel,
               },
               xmin=0.002,
               xmax=#5,
               ymin=0.002,
               ymax=#5,
               domain = 0.002:#5,
               grid=major,
              ]
            \addplot[marks] table [y index = 1, x index = 0, white space        chars={{,},\ }]  {#2}; 
            \addplot[draw=gray!50!white] {x};
        \end{axis}
    \end{tikzpicture}
}

\newcommand{\plotDataKilo}[5]{
    \begin{tikzpicture}
        \tiny
        \begin{axis}[height=0.35\textwidth, 
               width=0.35\textwidth,
               title={#1},
               xlabel={#3},
               ylabel={#4},
               every axis title/.append style={
                 at={(0.5,.95)}
               },
               every axis y label/.style={
                    at={(ticklabel cs:0.5)},rotate=90, anchor=near ticklabel,
               },
               every axis x label/.style={
                    at={(ticklabel cs:0.5)}, anchor=near ticklabel,
               },
               xmin=0,
               xmax=#5,
               ymin=0,
               ymax=#5,
               domain = 0:#5,
               grid=major,
               xtick={0, 5000,10000,15000},
               ytick={0, 5000,10000,15000,20000},
               xticklabels={0K, 5K, 10K, 15K},
               yticklabels={0K, 5K, 10K, 15K, 20K},
               xtick scale label code/.code={},
               ytick scale label code/.code={},
              ]
            \addplot[marks] table [y index = 1, x index = 0, white space        chars={{,},\ }]  {#2}; 
            \addplot[draw=gray!50!white] {x};
        \end{axis}
    \end{tikzpicture}
}

To analyze the practical aspects of finding minimal proofs, we ran several experiments for reasoning tasks in description logics.%
\footnote{The datasets and scripts used for the experiments can be found at \url{https://lat.inf.tu-dresden.de/2020-finding-small-proofs} .}
We first extracted a dataset containing explanation tasks from the ontologies of the OWL 2 EL classification track of the 2015 OWL Reasoner Evaluation (ORE)~\cite{ORE-2015}.
Due to the second experiment described below, we filtered out all sentences that are not expressible in \ELH.
To avoid re-evaluating proofs that are the same modulo renaming of concept and role names, we extracted a set of \emph{justification patterns}.
Specifically, we used \Elk to compute all justifications~$\Jmc$ of entailments~$\alpha$ of the form $A\sqsubseteq B$ or $A\equiv B$ in the filtered ontologies, and identified those that are the same modulo renaming, resulting in a set of 1573 unique entailments $\Jmc\models\alpha$ (up to renaming of concept and role names).%
\footnote{For $4$ ontologies the generation of justifications caused an OutOfMemoryError (using Java OpenJDK 11.0.4 with a maximum heap size of 23.5GB), hence any unique \ELH entailments from these ontologies are not part of our dataset.}

\subsection{Minimal Proofs vs.\ Minimal Tree Proofs}

\begin{figure}
  \centering
  \input{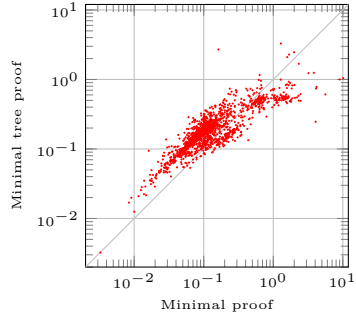}
  \caption{Times for extracting minimal (tree) proofs (in ms). Each dot represents one explanation task.}
  \label{fig:runtimes}
\end{figure}

Internally, \Elk generates a derivation structure that contains all possible proofs of a given entailment, possibly even including cycles. \Elk~0.5 provides an explanation service that can be used for directly accessing the proof steps~\cite{ELK-TRACING,KaKS-DL17}: specifically, for a given entailment $C\sqsubseteq D$, it allows to request all rule applications performed by \Elk that produce $C\sqsubseteq D$. Recursive requests of this kind allow to reconstruct the entire derivation structures generated by \Elk.
We used these derivation structures for our purposes to analyze the difference between \MP and \MTP in practice.
We implemented the Dijkstra-based algorithm for extracting minimal tree proofs, and a brute-force recursive procedure for finding minimal (non-tree) proofs.
The resulting runtimes for the 1573 explanation tasks are compared in Figure~\ref{fig:runtimes}.
One can see that extracting trees was generally slower, but gained an advantage for larger derivation structures.
This is probably due to the overhead of the various data structures required for the Dijkstra algorithm, the implementation of which was not optimized very far.

This gives experimental evidence of the difference in complexity of these two tasks observed in the previous section.
More surprisingly, however, it turns out that both methods always yielded equivalent results, \ie for our dataset the minimal tree proof is always the unraveling of the minimal (non-tree) proof.
Although in theory there can be a difference between the two, this does not seem to be relevant in practice, at least not in our dataset.

\subsection{Forgetting-Based Proofs}

Since finding a proof with a minimal number of vertices is \NP-hard, we now leave the goal of optimality aside and simply compare proofs generated by different techniques using different metrics.
Since there exist only few implementations of consequence-based algorithms for description logics that are able to produce proofs (\Elk~\cite{DBLP:journals/jar/KazakovKS14,KaKS-DL17} being the notable exception), we developed a black-box approach to generate proofs that are not based on a set of inference rules.

The main idea for this approach comes from the observation that the premises of a proof usually contain more symbols than the conclusion. For example, in the proof
\scalebox{0.65}{\AXC{$A\sqsubseteq C$}
\AXC{$C\sqsubseteq B$}
\BIC{$A\sqsubseteq B$}
\DP}
the concept name~$C$ is eliminated by the inference step.
Hence, the problem of finding a proof looks like a forgetting problem~\cite{FOUNDATIONS-UI,ALCH-FORGETTING}, where the symbols that do not occur in the conclusion should be removed.
In the example, $\{A\sqsubseteq B\}$ can be seen as the result of forgetting~$C$ in the original TBox $\{A\sqsubseteq C,\ C\sqsubseteq B\}$.
More generally, a \emph{forgetting-based proof} is composed of multiple steps that correspond to forgetting single symbols.
%

\begin{definition}
Given a TBox~\Tmc and a concept or role name~$X$, the result of \emph{forgetting~$X$ 
in~\Tmc} is another TBox~$\Tmc'$ such that $\sig{\Tmc'}\subseteq\sig{\Tmc}\setminus\{X\}$ 
and for any sentence~$\eta$ with $\sig{\eta}\subseteq\sig{\Tmc}\setminus\{X\}$ we have 
$\Tmc'\models\eta$ iff $\Tmc\models\eta$.
\end{definition}
The result of forgetting might not always exist in typical description logics, and require the use of 
additional constructs such as fixpoint 
operators~\cite{FOUNDATIONS-UI,ALCH-FORGETTING,FAME}. 

Our \emph{forgetting-based approach (FBA)} for constructing proofs can be summarized as follows. Given a TBox~\Tmc and a sentence~$\eta$ that is of the form $A\sqsubseteq B$ or $A\equiv B$, the goal is to forget all symbols except $A$ and $B$ from~\Tmc.
Since rarely the whole TBox is needed for a proof, we first extract a justification $\Jmc\subseteq\Tmc$ for~$\eta$, which already fixes the premises of the final proof.
We then construct a sequence of TBoxes $\Tmc_0,\Tmc_1,\dots,\Tmc_n$, where 
$\Tmc_0=\Jmc$ and each $\Tmc_i$, $1\le i\le n$, is the result of forgetting one symbol 
from~$\Tmc_{i-1}$ (except~$A$ or~$B$), and then extracting a justification for~$\eta$. This 
process finishes when $\sig{\Tmc_n}\subseteq\{A,B\}$.
If one forgetting step fails, we try to forget a different symbol instead.
If this fails for all symbols in the current signature, the process terminates early and we set $\Tmc_n:=\{\eta\}$.

To reconstruct the actual proof from these TBoxes, we start with a vertex labeled with $\eta$ and select a justification~$\Jmc_\eta$ for~$\eta$ in~$\Tmc_{n-1}$, which gives us a first proof step that derives~$\eta$ from~$\Jmc_\eta$.
We add a new vertex for each element of $\Jmc_\eta$, and a hyperedge connecting these vertices to the sink.
We then recursively justify each element of~$\Jmc_\eta$ in~$\Tmc_{n-2}$, and continue in this manner until we reach the sentences from the original TBox in $\Tmc_0=\Jmc\subseteq\Tmc$.
Each justification step for an intermediate sentence then corresponds to one inference step in the proof, which can be further annotated by the symbol that was forgotten in the corresponding forgetting operation.

\begin{example}
\label{ex:906}
We start with $\Tmc_0=\{C\sqsubseteq\exists r.D,\ A\sqsubseteq C,\ \exists r.\top\sqsubseteq B\}$ and want to prove that $\Tmc_0\models A\sqsubseteq B$.
We first forget the symbol~$D$, resulting in $\Tmc_1=\{C\sqsubseteq\exists r.\top,\ A\sqsubseteq C,\ \exists r.\top\sqsubseteq B\}$, then $C$, which yields $\Tmc_2=\{A\sqsubseteq\exists r.\top$, $\exists r.\top\sqsubseteq B\}$.
Finally, forgetting~$r$ obtains $\Tmc_3=\{A\sqsubseteq B\}$; see Figure~\ref{fig:FBAexample}.
%
\begin{figure}[tb]
\centering
\begin{minipage}[b][][b]{0.495\textwidth}
\begin{prooftree}
\AXC{$A\sqsubseteq C$}
\AXC{$C\sqsubseteq\exists r.D$}
\LeftLabel{\upshape($D$)}
\UIC{$C\sqsubseteq\exists r.\top$}
\LeftLabel{\upshape($C$)}
\BIC{$A\sqsubseteq\exists r.\top$}
\AXC{$\exists r.\top\sqsubseteq B$}
\LeftLabel{\upshape($r$)}
\BIC{$A\sqsubseteq B$}
\end{prooftree}
\caption{A proof obtained by FBA}\label{fig:FBAexample}
\end{minipage}
\hfill
\begin{minipage}[b][][b]{0.495\textwidth}
\raggedleft
\begin{prooftree}
\AXC{$A\sqsubseteq C$}
\AXC{$C\sqsubseteq\exists r.D$}
\AXC{$\exists r.\top\sqsubseteq B$}
\LeftLabel{\upshape($r$)}
\BIC{$C\sqsubseteq B$}
\LeftLabel{\upshape($C$)}
\BIC{$A\sqsubseteq B$}
\end{prooftree}
\caption{An alternative proof}\label{fig:FBAexample-other}
\end{minipage}
\end{figure}
The premises of \eg $A\sqsubseteq\exists r.\top$ in~$\Tmc_2$ are a justification of this sentence in~$\Tmc_1$.
\end{example}

Note that the precise result of FBA depends on the choices of justifications in each step as well as the order in which symbols are forgotten.

\begin{example}
First forgetting~$r$ in~$\Tmc_0$ from Example~\ref{ex:906} would result in $\Tmc_1=\{A\sqsubseteq C,\ C\sqsubseteq B\}$ and a different proof  in Figure~\ref{fig:FBAexample-other}.
\end{example}

However, regardless of these choices, FBA is sound and complete.
\begin{restatable}{theorem}{FBA}\label{th:forgetting-based-proof}
Let FBA use a sound and complete reasoner to compute justifications and a sound forgetting tool. Given a sentence $\eta$ and a TBox \Tmc, FBA produces a proof \p iff $\Tmc\models \eta$.
\end{restatable}

The main advantage of this approach is that it works in a black-box fashion, using any forgetting tool and reasoner to compute justifications (as long as they support \L).
This means that we can find proofs for inferences in expressive logics like \ALCH or \ALCOI, for which no consequence-based proof generators exist, using forgetting tools such as \Lethe~\cite{LETHE} and \Fame~\cite{FAME}.
FBA also does not explicitly generate tautologies, as many consequence-based reasoners do, cutting down on the size of proofs, but possibly making them harder to follow.
An obvious disadvantage compared to consequence-based reasoners is the lack of a predefined set of inference rules, requiring more effort to understand each individual inference step.

\subsection{Optimizations for FBA}

When actually implementing FBA, we considered several refinements of this method.
First, it may be the case that the justification $\Jmc_\alpha$ for $\alpha\in\Tmc_i$ in~$\Tmc_{i-1}$ may be exactly $\{\alpha\}$, because~$\alpha$ does not contain the symbol that was forgotten in this step.
In such a case, we skip the corresponding trivial inference step
\scalebox{0.65}{\AXC{$\alpha$}\UIC{$\alpha$}\DP},
and instead use a justification for~$\alpha$ from the first TBox~$\Tmc_j$, $j<i$, such that $\alpha\notin\Tmc_j$.
We further extended this optimization: if the justification of a sentence $\alpha\in\Tmc_i$ in $\Tmc_{i-2}$ does not contain more sentences than the justification in $\Tmc_{i-1}$, we choose the premises from the \enquote{earlier} TBox~$\Tmc_{i-2}$ to skip \enquote{simple} steps involving only one sentence.
In Example~\ref{ex:906}, the justification $\{A\sqsubseteq C,\ C\sqsubseteq\exists r.D\}$ for $A\sqsubseteq\exists r.\top$ in~$\Tmc_0$ would be preferred to the justification of the same size in~$\Tmc_1$, resulting in the more compact proof -- see Figure~\ref{fig:FBAoptimisation}.
\begin{figure}[tb]
\begin{prooftree}
\AXC{$A\sqsubseteq C$}
\AXC{$C\sqsubseteq\exists r.D$}
\LeftLabel{\upshape($C$)}
\BIC{$A\sqsubseteq\exists r.\top$}
\AXC{$\exists r.\top\sqsubseteq B$}
\LeftLabel{\upshape($r$)}
\BIC{$A\sqsubseteq B$}
\end{prooftree}
\caption{A more compact proof for Example~\ref{ex:906}}
\label{fig:FBAoptimisation}
\end{figure}

Regarding the order in which symbols are forgotten, we considered several aspects.
Once a role name is forgotten, all proof steps that concern the fillers of existential or value restrictions over this role are hidden from the proof. This is problematic if most proof steps take place \enquote{within} a role restriction, which is why role names should be forgotten as late as possible to make the proofs as fine-grained as possible.
On the other hand, once a role~$r$ does not have non-trivial fillers anymore, \ie only occurs in expressions like $\exists r.\top$ or $\forall r.\bot$, we should immediately forget it, to focus on the remaining parts of the sentence instead of keeping $\exists r.\top$ around in all intermediate steps.
More precisely, the priority with which we try to forget symbols in each step is the following: first role names without non-trivial fillers, then concept names that do not occur nested within existential or value restrictions, then any other concept names, and finally the remaining role names.


\subsection{Evaluating FBA}

%
In our second experiment, we compared the minimal proofs extracted from \Elk with the ones from our forgetting-based approach, using both \Lethe~\cite{LETHE} and \Fame~\cite{FAME} as black-box forgetters%
\footnote{Both \Lethe and \Fame may generate fixpoint expressions that are not expressible in usual DLs. We omit any such steps from the generated proofs.}
(and HermiT~\cite{HERMIT} as reasoner).
Of \Lethe, we used the latest version 0.6,\footnote{\url{https://lat.inf.tu-dresden.de/~koopmann/LETHE}}
and for \Fame, we used the \ALCOI-variant of \Fame~1.0,%
\footnote{\url{http://www.cs.man.ac.uk/~schmidt/sf-fame}} since other versions often 
fail to produce forgetting results that can be handled by the OWL API. Apart from the supported 
DL, a major difference between \Lethe and \Fame is that \Lethe is theoretically guaranteed to 
compute a sound and complete forgetting result, while \Fame is incomplete and may fail to 
forget a given name.
Note that \Lethe supports \ALCH and \Fame supports \ALCOI, but we only consider \ELH reasoning tasks since we want to compare with \Elk.

To demonstrate the output of the proof generators, in Figure~\ref{fig:ex4} we provide two alternative proofs
, one generated by \Elk and the other by FBA.
The automatically generated images use blue boxes to denote hyperedges $(S,d)$, with incoming and outgoing arrows 
indicating~$S$ and~$d$, respectively, and a label indicating the type of inference rule (for \Elk) 
or the forgotten symbol(s) (for FBA).
In contrast to our previous definition of hypergraphs, both \Elk and our implementation of FBA treat sentences from the TBox not as leaves, but as being derived by a special kind of hyperedges $(\emptyset,v)$ labeled by ``asserted'' or ``Asserted Conclusion''.

\iftechnicalReport
\begin{figure}
  \centering
  \includegraphics[width=.32\linewidth]{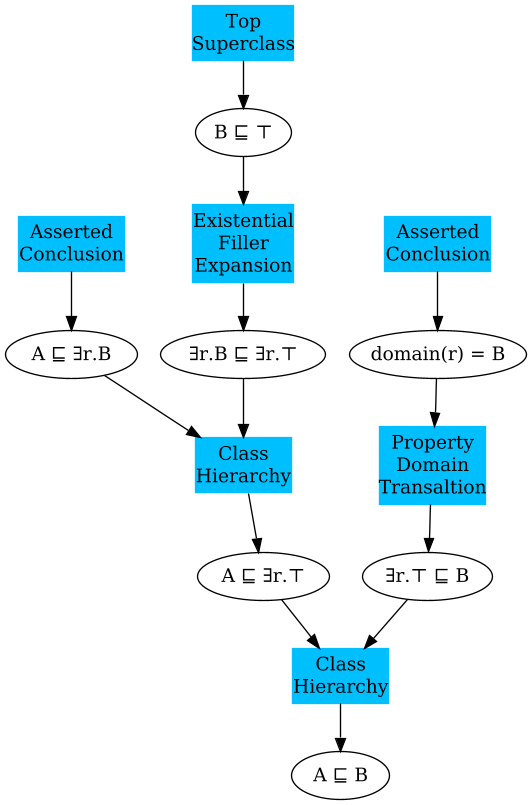}
  \qquad
  \includegraphics[width=.21\linewidth]{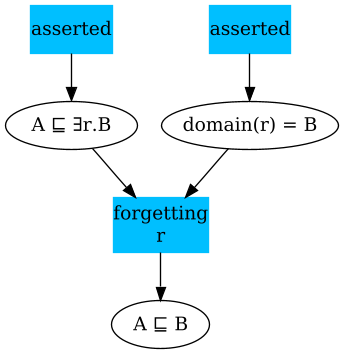}
  \caption{The \Elk proof (left) makes several inference steps, which are summarized by \Lethe and \Fame (right) into a single step.}
  \label{fig:ex4}
\end{figure}
\fi

\begin{figure}[tb]
\input{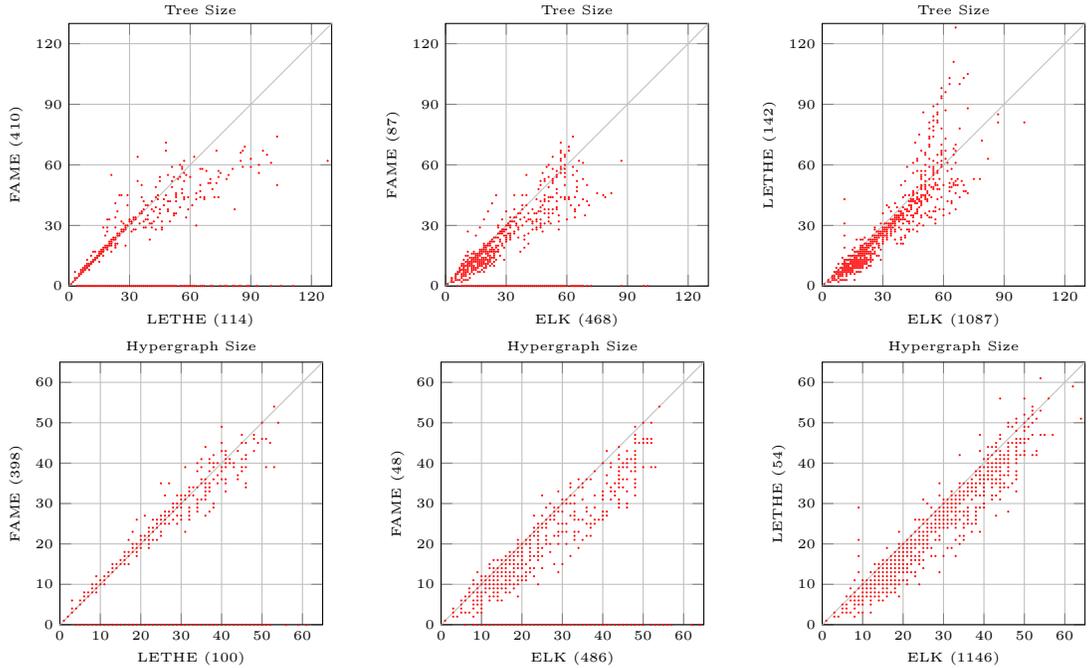}
\caption{Comparison of the tree size and hypergraph size of the proofs generated by the three tools. In parentheses are the numbers of entries that lie strictly above or below the diagonals (excluding cases where one of the tools did not produce a proof at all, which are evaluated to~$0$).}
\label{fig:tree-size}
\end{figure}

We generated such proofs for each explanation task from our dataset using the 3 tools and compared these proofs according to different metrics.
In Figure~\ref{fig:tree-size} we compare the \emph{hypergraph size} (number of vertices) and \emph{tree size} (number of vertices of the unraveling) of the generated proofs, where each dot may represent multiple entailments $\Jmc\models\alpha$. \Fame did not generate proofs for 656 of the 1573 tasks, which are assigned a score of~$0$ in all diagrams (290 contained role inclusions, which are not supported by the version of \Fame that we used, and in 366 cases it returned an incorrect forgetting result\footnote{As this points at a bug in the tool that might not have been detected before, we have informed the developer.}).
In contrast, \Lethe is complete for \ALCH forgetting and always yields a proof using the preferred order of forgetting symbols.
We can see that most of the forgetting-based proofs were of similar size or smaller than the corresponding ones generated by \Elk.
However, recall that the proofs generated using \Lethe and \Fame have the advantage that they can use inference steps in more expressive logics than \ELH, which may result in more compact proofs.
Moreover, the proofs generated using \Fame were usually larger than the ones from \Lethe, 
which can be explained by additional simplification steps performed by \Lethe when computing 
the forgetting result~\cite{LETHE}.

\begin{figure}[tb]
\input{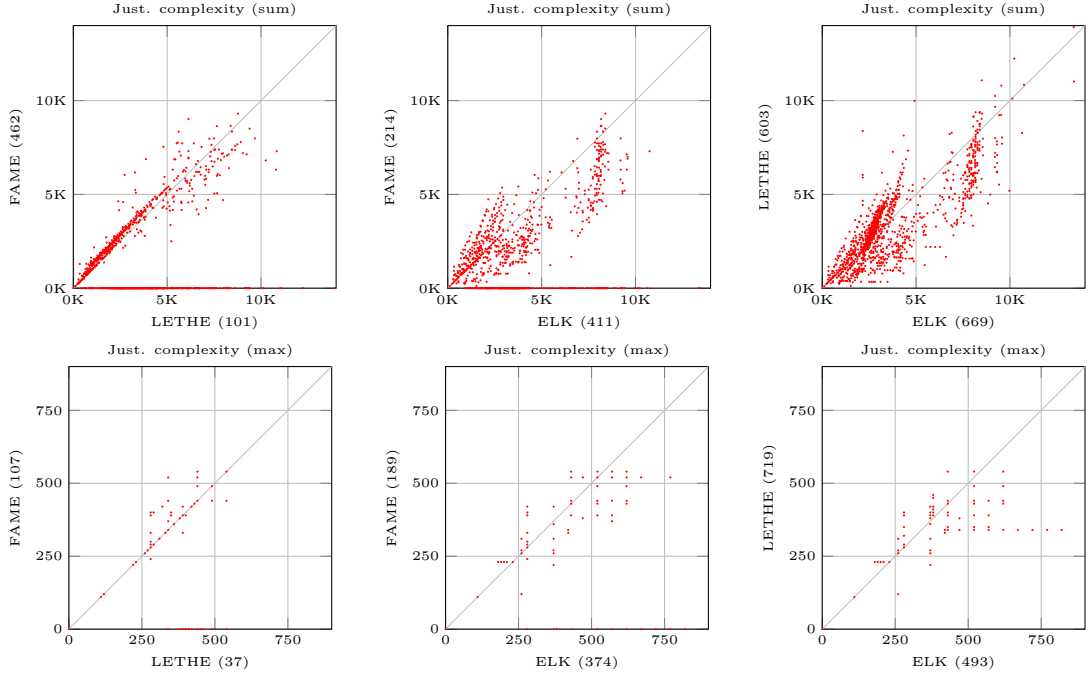}
\caption{The total and maximal justification complexity~\cite{HBPS-KBS13} of proofs.}
\label{fig:just-comp}
\end{figure}

Next we wanted to find out how hard it would be for a user to actually understand the generated proofs. This is hard to evaluate empirically, but was already attempted for plain justifications~\cite{HBPS-KBS13}, resulting in a \emph{justification complexity model} that assigns a numerical score to an entailment task $\Jmc\models\alpha$, indicating its difficulty.
We evaluated all individual inference steps in the generated proofs according to this model and report in Figure~\ref{fig:just-comp} the sum and the maximum of the justification complexities of all inference steps in each proof.
Here, the advantage of the forgetting-based approach over \Elk was less pronounced.
Although the sum of justification complexities is correlated with the hypergraph size, the difficulty of the individual steps now also plays a role.
For maximum justification complexity, \Lethe scored worse than \Elk more often than vice versa, but for \Fame and \Elk the situation is reversed (on a restricted dataset, because \Fame did not generate proofs in all cases).

Looking at individual examples, however, the simplicity of the \Elk inference rules is clearly an advantage when one wants to understand the generated proofs.
In total, we could find 362 distinct \enquote{inference rules} used in the proofs generated using \Lethe (up to renaming of concept and role names).
For \Fame, we extracted 281 inference rules.
This large diversity is obviously not captured by the hypergraph size, but also not accurately represented by the justification complexity, which only measures single inference steps.
Moreover, understanding inference steps formulated in \ALCH or \ALCOI is inherently harder than for pure \ELH proofs.
%
In the \iftechnicalReport appendix\fi, we show some more examples of the generated proofs.

We also attempted some experiments on \ALCH ontologies from the ORE dataset, but were not able to extract many explanation tasks (that were not already in the \ELH dataset), mostly due to time and memory problems when computing all possible justifications for all \ALCH entailments using a black-box approach based on HermiT (for the \ELH dataset we could directly use \Elk to compute all justifications).


\section{Conclusion}

We have investigated the complexity as well as the practical aspects of finding small proofs for description logic entailments.
Obviously, size is not enough to evaluate the difficulties users face when trying to understand proofs, and even existing measures such as the justification complexity model from~\cite{HBPS-KBS13} cannot capture all possible aspects of this problem.
Hence, in future work we will extend our results to more complex measures of understandability for proofs.
We conjecture that many of our results can be transferred, \eg when the goal is to minimize the sum of justification complexities we also obtain an \NP-complete decision problem, whereas for the maximum of justification complexities we expect a \PTime-upper bound since such maxima can also be computed by a greedy, Dijkstra-like approach.

\paragraph{Acknowledgements} This work was partially supported by DFG grant 389792660 
as part of TRR~248 (\url{https://perspicuous-computing.science}), and the DFG Research 
Training Group QuantLA, GRK 1763 (\url{https://lat.inf.tu-dresden.de/quantla}).

\bibliographystyle{plainurl}
\bibliography{bibs}

\iftechnicalReport
\setcounter{section}{0}
\renewcommand{\thesection}{\Alph{section}}

\clearpage

\section{Proofs Omitted from the Main Text}\label{appendix:proofs}
Since we investigative the problem of finding a minimal proof, we concentrate on proofs without redundant elements. 
\begin{definition}[Redundant]
We call a proof \p for~$\Tmc\models\eta$ \emph{redundant} if it properly contains another proof for~$\Tmc\models\eta$; otherwise it is \emph{non-redundant}.
\end{definition}
A non-redundant proof for~$\Tmc\models\eta$ cannot contain several vertices labeled with~$\eta$, or two inference steps that lead to the same vertex (proved by Lemma~\ref{lem:proof-properties}\,(3) below).
However, it is still possible that there are two vertices labelled with the same sentence.

For proofs we can show some important properties.
\begin{lemma}\label{lem:proof-properties}
Let $\p=(V,E,\el)$ be a proof for~$\Tmc\models\eta$. Then
\begin{enumerate}
\item all paths in \p are finite and all longest paths in \p have $v_\eta$ as the target;
\item $\Tmc\models \eta$;
\item if $\p$ is non-redundant, there is no vertex $w\in V$ \st there are $(S_1,w), (S_2,w)\in E$ with $S_1\neq S_2$.
\end{enumerate}
\end{lemma}
\begin{proof}
The first statement trivially follows from the acyclicity and the only sink $v_\eta$ in \p. The length of a path in \p can be bounded by~$|V|$.

The second claim can be shown by an induction on the depth of \p. Namely, for every $k$ and every $w\in V$ \st all paths leading to~$w$ have length at most~$k$ it holds that $\Tmc\models \el(w)$. The induction base follows from the fact that the leafs are labelled with the sentences from \Tmc. For the induction step, for a vertex $w\in V$, we consider an hyperedge $(S,w)\in E$. Every $s\in S$ satisfies the induction hypothesis and, thus,  $\Tmc\models \el(s)$. By \p being a derivation structure, it holds $\{\el(s) | s\in S\} \models \el(w)$ and, by transitivity of model-based entailment, $\Tmc\models \el(w)$.  

Now, we consider a non-redundant \p. Assume that there is a vertex $w \in V$ with two incoming edges, $(S_1,w)$ and $(S_2,w)$. By removing one of the edges, say $(S_1,w)$, and, if a vertex $v\in S_1$ becomes a sink, removing it and all its incoming edges $(S,v)$, and continuing recursively in this fashion, we obtain a proper substructure of~\p. The new derivation structure still has the original sink~$v_\eta$ from~\p and no cycles can be created by the described procedure. This contradicts \p being non-redundant. 
\end{proof}

Since redundant proofs will in general not be minimal, we can restrict most of our considerations to non-redundant proofs.

\hyperproofInside*
\begin{proof}
Let $\p=(V,E,\el)$ be a minimal proof \wrt $\R(\Tmc,\eta)$.
By definition, there is a homomorphism $h\colon \p \rightarrow \R(\Tmc,\eta)$.
We show that $h$ is injective.
Assume to the contrary that there are two vertices $v,v'\in V$ with $h(v)=h(v')$, which implies that $\el(v)=\el(v')$.
If there is no path between $v$ and $v'$ in~\p, then we can construct~$\p'$ by renaming all occurrences of~$v$ in~\p to~$v'$.
This operation does not introduce any leafs, sinks, or cycles (in particular no cycles involving the sink of~\p), and hence $\p'$ is still a proof for~$\Tmc\models\eta$.
But $\ct{\p'}<\ct{\p}$, which contradicts minimality of~\p.

It remains to deal with the case where there is a path between~$v$ and $v'$.
In this case, we ensure that $v$ and $v'$ are not connected anymore before merging them.
Since \p is acyclic, we can assume that there is a path from~$v$ to~$v'$, but no path from~$v'$ to~$v$.
We now start from~$v'$ and remove~$v'$ and all hyperedges of the form $(S,v')$ from~$\p$, which already destroys any paths from~$v$ to~$v'$.
However, this operation may make some vertices $w\in S$ into sinks, and we continue removing such~$w$ and hyperedges $(S',w)$ until either no new sinks are produced or we have reached~$v$.
The only remaining loose ends are the hyperedges $(S,d)$ in~\p that still contain $v'\in S$.
In such hyperedges, we replace $v'$ by~$v$, effectively merging the two vertices.
If there was no such hyperedge, then $v'$ was the sink of \p, \ie $\el(v')=\eta$, and $v$ will now be the new sink with $\el(v)=\el(v')=\eta$.

We show that the result~$\p'$ of this procedure is still a proof, which has less vertices than~\p, which again contradicts our assumption on~\p.
Since our procedure removes edges only together with their destination vertices, it cannot produce new leafs, and hence all leafs are still labeled by sentences from~\Tmc.
Clearly, all remaining edges are sound since they were already sound in~\p.
As argued above, we have also kept the property that there is exactly one sink, which is labeled with~$\eta$.
Finally, all cycles in~$\p'$ can be traced back to paths in~\p that involve both~$v$ and~$v'$.
But we have assumed that there are no paths from~$v'$ to~$v$, and have destroyed all paths from~$v$ to~$v'$.

This shows that $h$ must be injective, and therefore its image $h(\p)$ is isomorphic to~$\p$.
This means that $h(\p)$ is a substructure of $\R(\Tmc,\eta)$ which is a proof for~$\Tmc\models\eta$ that is also minimal because $\ct{h(\p)}=\ct{\p}$.
%
\end{proof}
%

We first show the upper bounds for the second row of Table~\ref{tab:results1}.

\thPolyNPExpNExpMem*
\begin{proof}
  We start by guessing a substructure~\p of $\R(\Tmc,\eta)$.
  This can be done in polynomial time in the size of $\R(\Tmc,\eta)$, which for $\MP^\poly_\bin$ is polynomial in the size of~$\Tmc$ and~$\eta$, and for  $\MP^\exp_\bin$ is exponential.
  This substructure~\p is necessarily sound and checking the remaining conditions of Definitions~\ref{def:derivation-structure} and~\ref{def:proof} and that \p is not larger than $n$ can be done in polynomial time in the number of vertices in~\p.
  %
\end{proof}

In the case of~$\MP^\exp_\un(\R)$, we can also show an \NP-upper bound, but for this we have to be more careful about how we access the structure $\R(\Tmc,\eta)$, which is already of exponential size.

Namely, we make a few additional assumptions about how the \reasoner{} works. These assumptions do not make a difference for the other cases, but enable us to actually guess a structure of polynomial size (at most~$n$ vertices and one edge for each vertex) without needing to store $\R(\Tmc,\eta)$, which is of exponential size.
  
  Let $\ds=(V,E,\el):=\R(\Tmc,\eta)$.
  We assume that the vertices $V=\{v_1,\dots,v_m\}$ are numbered consecutively, where $m$ is an index of polynomial size (in binary notation).
  Hence, in order to address a single vertex $v_i$, we only need access to its index $i$, $1\le i\le m$, instead of to the whole set~$V$.
  Furthermore, we only access the structure~\ds by means of two oracle functions
  \begin{align*}
    [\ds](v_1,\dots,v_l) &:=
    \begin{cases}
      \true &\text{if $(\{v_1,\dots,v_{l-1}\},v_l)\in E$,} \\
      \false &\text{otherwise;}
    \end{cases} \\
    [\ds](v,\alpha) &:=
    \begin{cases}
      \true & \text{if $\el(v)=\alpha$,} \\
      \false & \text{otherwise.}
    \end{cases}
  \end{align*}
  where $v,v_1,\dots,v_l$ are vertices (represented by their indices) and $\alpha$ is an sentence.
  These functions hide any details of the internal representation of $\ds$.
  For example, if $v_1,\dots,v_m$ is an efficient indexing scheme for the (exponentially many) sentences in~\ds, then the call to $[\ds](v_1,\dots,v_l)$ can be answered by retrieving the sentences $\alpha_1,\dots,\alpha_l$ corresponding to $v_1,\dots,v_l$ in polynomial time and checking whether {\small\insertBetweenHyps{\hskip0em}\AXC{$\alpha_1$}
\AXC{$\ldots$}\AXC{$\alpha_{l-1}$}\TIC{$\alpha_l$}\DP} is a valid rule instance.

Having fixed a theory \Tmc and a sentence $\eta$, by this restricted view on the exponentially large output of a \reasoner{}, we can show the following result.
\thExpUnaryNPMem*
\begin{proof}
Using the setup described above, we can guess $n'\le n$ vertices $v_{i_1},\dots,v_{i_{n'}}$ and for each vertex at most one incoming edge $(S_{i_j},v_{i_j})$ with $S_{i_j}\subseteq\{v_{i_1},\dots,v_{i_{n'}}\}$, in polynomial time in the sizes of~$n$, \Tmc, and~$\eta$.
  Recall that we can restrict our search to non-redundant proofs here since we are looking for minimal proofs anyway.
  We can then check the following conditions to ensure that we have guessed a valid proof.
  
  First, we make sure that every guessed edge $(\{v_1,\dots,v_{l-1}\},v_l)$ is actually an edge in~\ds by calling $[\ds](v_1,\dots,v_l)$; this also guarantees soundness of the proof since \ds is a derivation structure.
  To check completeness, we use $[\ds](v,\alpha)$ to ensure that any vertex~$v$ for which we have not guessed an incoming edge is labeled by some $\alpha\in\Tmc$.
  Finally, we use several reachability checks in the guessed structure to make sure that there is a unique sink, which is labeled by~$\eta$, and that there are no cycles.
  %
\end{proof}

As a consequence of the theorem above, $\MP^\poly_\un(\R)$ is also in~\NP. Moreover, the \NP-hardness can be shown by a reduction of the hitting set problem~\cite{10.5555/574848}, which is inspired by a proof about minimal justifications in~\cite{DBLP:conf/ki/BaaderPS07}.
\thPolyGraphNPhard*
\begin{proof}
  Given a finite set $\universe = \{ u_1, \ldots, u_l \}$, sets $s_i \subseteq \universe$, $1 \leq i \leq k$, and a number $m\in\nats$, we need to decide whether there is a set $s\subseteq\universe$ of size less than~$m$ such that $s_i \cap s \neq \emptyset$, $1\le i\le k$.
  We use an instance of $\MP^\poly_\un(\text{ELK})$ with an \EL-TBox~\Tmc, sentence ~$\eta$ and number~$n$ as follows.

  Let $A,B,U_1,\dots,U_l,S_1,\dots, S_k$ be concept names. \Tmc contains the following sentences:
  \begin{align*}
  A &\sqsubseteq U_i, && \text{ for every $i$, } 1\leq i \leq l; \\
  U_i &\sqsubseteq S_j, && \text{ for every }  u_i\in s_j, 1\leq j \leq k, 1\leq i\leq l; \\
  S_1\sqcap\dots\sqcap S_k &\sqsubseteq B. & 
  \end{align*}
  A proof of $\eta:=A\sqsubseteq B$ over~\Tmc using the inference rules of ELK~\cite{DBLP:journals/jar/KazakovKS14} first derives $A\sqsubseteq S_j$ for each $j$, $1\le j\le k$, from $A\sqsubseteq U_i$ and $U_i\sqsubseteq S_j$ by rule $\mathsf{R_\sqsubseteq}$ (see Figure~\ref{fig:cr}) for some $i$, $1\le i\le l$, where the same sentence $A\sqsubseteq U_i$ can be used for different~$S_j$.
  Then it composes the sentences $A\sqsubseteq S_j$ into $A\sqsubseteq S_1\sqcap\dots\sqcap S_k$ with $\mathsf{R_\sqcap^+}$ using $k-1$ inference steps, before finally deriving $A\sqsubseteq B$ with the help of $S_1\sqcap\dots\sqcap S_k \sqsubseteq B$ and $\mathsf{R_\sqsubseteq}$.
  The number of sentences in this proof is $m'+3k+1$, where $m'$ is the number of sentences $A\sqsubseteq U_i$, which represent the hitting set~$s$.
  Hence, there is a hitting set of size $\le m$ iff there a proof as described above with $\le n := m+3k+1$ vertices.
\end{proof}

The result in Theorem~\ref{th:poly-graph-NPhard} also provides the lower bound for all problems in Table~\ref{tab:results1}. However, for an exponential \reasoner{} and binary encoding, we are able to strengthen it to \NExpTime.

\begin{figure}[tb]
  \input{figures/eli-calculus}
  \caption{The inference rules for \ELI~\cite{baader_horrocks_lutz_sattler_2017}. Given a TBox $\Tmc$ in a normal form, the rules produce a saturated TBox~$\Tmc'$. Here, $K,L,M$ are conjunctions of concept names, $A$ is a concept name, $C$ is an \ELI concept of the form $A$, $\exists r.M$, or $\forall r. A$, and $r$ is a role name or the inverse of a role name.
  In this calculus conjunctions are implicitly viewed as sets, \ie the order and multiplicity of conjuncts is ignored.}
  \label{fig:ELIcr}
\end{figure}


We first state a result on tiling problems to simplify the proof of the theorem. We consider the following \emph{exponential tiling problem}: given a set $T$ of tiles, two sets $h,v\subseteq T\times T$ of horizontal and vertical tiling conditions, an initial sequence $t_0,\ldots,t_s\in T$ of tiles, and a number~$n$ encoded in unary, decide whether there exists a function $f\colon[0,2^n-1]\times[0,2^n-1]\rightarrow T$ such that $f(i,0)=t_i$ for $0\leq i\leq s$, $\tup{f(i,j),f(i+1,j)}\in h$ and $\tup{f(j,i),f(j,i+1)}\in v$ for all $i\in[0,2^n-2]$ and $j\in[0,2^n-1]$. This problem is known to be \NExpTime-complete~\cite{TILINGS}. Furthermore, it is sufficient to consider a variant for which $s=0$.

\begin{lemma}
 Let $T$, $h,v\subseteq T\times T$, $n$ and $t_0,\ldots,t_s\in T$ be a tiling problem. Then, there exists a tiling problem $T'$, $h',v'\subseteq T'\times T'$ with just one initial tile $t_0'\in T$ that has a solution iff the original tiling problem has a solution.
\end{lemma}
\begin{proof}
 We assign to every tile $t_i$, $1\leq i\leq s$, a unique fresh tile $t_i'$, and then define the new tiling problem as follows.
 \begin{align*}
 h' = h &\cup \{\tup{t_0',t_1'},\ldots,\tup{t_{s-1}',t_s'}\} \\
        &\cup \{\tup{t_s',t}\mid \tup{t_s,t}\in h\} \\
 v' = v &\cup \{\tup{t_i',t}\mid 0\leq i\leq s, \tup{t_i,t}\in v\}.
 \end{align*}
 It is easy to verify that this tiling problem satisfies the requirements.
\end{proof}

\thExpGraphNExpHard*
\begin{proof}
  We use the \reasoner generated by the calculus for \ELI shown in Figure~\ref{fig:ELIcr}, assuming that no sentence labels more than one vertex. The calculus is taken from~\cite{baader_horrocks_lutz_sattler_2017} with small modifications:
  \begin{itemize}
    \item While the authors assume the input to use at most two conjuncts on the left-hand side, we do not impose such a restriction. It is not hard to see that lifting this restriction neither affects soundness nor completeness of the method, and does not have an impact on the complexity of the problem considered.
    \item The authors of~\cite{baader_horrocks_lutz_sattler_2017} use a special set notation to represent $n$-ary conjunctions in inferred sentences. We just use $n$-ary conjunctions instead of sets, and assume that those are treated silently as sets when rules are applied: that is, conjunctions never have duplicate conjuncts and the order is not important.
  \end{itemize}
  
  To show \NExpTime-hardness for this \reasoner, we describe a reduction from the exponential tiling problem with one initial tile. In the following, let $T$, $h,v\in T\times T$ and $t_0\in T$ be such a tiling problem. For $i$, $1\leq i\leq \lvert h\rvert$, let $h_i$ denote the $i$th element of $h$, and for $i$, $1\leq i\leq \lvert v\rvert$, let $v_i$ denote the $i$th element of $v$. Furthermore, let $T_I\subseteq T$ be the set of tiles that occur at least once as first component in $h$ or $v$. 
  
  
  \newcommand{\Start}{\textsf{Start}\xspace}
  \newcommand{\Finish}{\textsf{Finish}\xspace}
  \newcommand{\Success}{\textsf{Success}\xspace}

  For the first reduction, consider the number of sentences $m:=m_1+m_2+m_3+m4$, which will be the size of the proof, where 
  \begin{align*}
  m_1 &:= 4n^2+6n+|h|+|v|+7, \\
  m_2 &:= (2^n-1)^2(10n+3)+2(2^n-1)(6n+2)-(2n+1),\\
  m_3 &:= 6(2^n-1)+5(2^n-1)^2+1,\text{ and }\\ 
  m_4 &:= \lvert T_I\rvert + 7\cdot (\lvert v\rvert + \lvert h\rvert) + 2,
  \end{align*}
  which can be computed in polynomial time when encoded in binary, and the TBox~$\Tmc=\Tmc_1\cup\Tmc_2$ shown in Figure~\ref{fig:nexptime-reduction-tbox} and~\ref{fig:nexptime-reduction-tbox2}. In the reduction, we enforce the grid-shape of the tiling using the size restriction of the proof. This means that we have to be very careful with the use of auxiliary sentences: if those are not needed in every proof, this could introduce degrees of freedom in other parts of the proof, so that we fail to enforce this grid-shape. Therefore, we have to make sure that the number of auxiliary sentences used is always the same independent of the tiling solution. Specifically, not all of the vertical or horizontal tiling conditions might be relevant for a tiling solution, but the axioms used to encode them must be used in every proof of our entailment. 
  $\Tmc_1$ in Figure~\ref{fig:nexptime-reduction-tbox} contains the main reduction of the tiling problem, while $\Tmc_2$ in Figure~\ref{fig:nexptime-reduction-tbox} makes sure that all of the sentences used to encode the tiling conditions are necessary for the proof.

  We use concept names $A_i$, $\overline{A}_i$, $B_i$, $\overline{B}_i$ $1\leq i\leq n$, to encode two binary counters for the $x$- and the $y$-coordinates in the tiling, respectively, where $A_i$ denotes that the $i$-th bit has a value of~$1$, and $\overline{A}_i$ that the $i$-th bit has a value of~$0$, and correspondingly for~$B$. For a number $i\in[0,2^n-1]$, we use $[A=i]$ as an abbreviation of the conjunction of concepts $A_i$, $\overline{A}_i$ that corresponds to a counter value of~$i$, and correspondingly for $[B=i]$.
  %
  We prove that $\Tmc\models\Start\sqsubseteq\Success$ has a proof using $\le m$ sentences iff the given tiling problem has a condition-complete solution.
  
  \begin{figure}
   \input{figures/nexptime-reduction-tbox}
   \caption{The TBox $\Tmc_1$ used in the reduction for Theorem~\ref{th:poly-graph-NPhard}.}
   \label{fig:nexptime-reduction-tbox}
  \end{figure}

  $(\Leftarrow$): Assume that the tiling problem has a solution~$f$. We describe how a proof of $\Tmc\models\Start\sqsubseteq\Success$ can be constructed.  %
  In this proof, we use concepts of the form 
  $$C(i,j):=f(i,j)\sqcap[A=i]\sqcap[B=j]$$ 
  on the left-hand side of GCIs to identify individual grid cells~$(i,j)$, and the right-hand side of such GCIs to derive properties of these cells.
  \begin{figure}
      \centering
    \begin{tikzpicture}[on grid,node distance=21ex and 8.5em,label distance=1ex,cell/.style={font=\tiny,draw=gray,text width=6.5em,align=center,minimum height=18ex},>=latex,shorten >=1pt,shorten <=1pt]
      \node[cell] (start) {
        $\exists r_x.f(1,0)$ $\exists r_y.f(0,1)$
        $\forall r_x.A_1 \dots \forall r_x.\overline{A_n}$
        $\forall r_x.\overline{B_1} \dots \forall r_x.\overline{B_n}$
        $\forall r_y.\overline{A_1} \dots \forall r_x.\overline{A_n}$
        $\forall r_y.B_1 \dots \forall r_x.\overline{B_n}$
        \vphantom{$r_x^-$}$\Finish_x$ $\Finish_y$ $\Finish$
      };
      \node[cell,above=of start] (above) {
        $\exists r_x.f(1,1)$ $\exists r_y.f(0,2)$
        $\forall r_x.A_1 \dots \forall r_x.\overline{A_n}$
        $\forall r_x.B_1 \dots \forall r_x.\overline{B_n}$
        $\forall r_y.\overline{A_1} \dots \forall r_x.\overline{A_n}$
        $\forall r_y.B_1 \dots \forall r_x.\overline{B_n}$
        \vphantom{$r_x^-$}$\Finish_x$ $\Finish_y$ $\Finish$\\
        $\forall r_y^-.\Finish_y$
      };
      \node[cell,draw=none,above=of above] (above2) {$\vdots$};
      \node[cell,above=of above2] (topleft) {
        $\exists r_x.f(1,2^n{-}1)$
        $\forall r_x.A_1 \dots \forall r_x.\overline{A_n}$
        $\forall r_x.B_1 \dots \forall r_x.B_n$
        \vphantom{$r_x^-$}$\Finish_x$ $\Finish$
        $\forall r_y^-.\Finish_y$
      };
      \node[cell,draw=none,right=of above] (diag) {\rotatebox{90}{$\ddots$}};
      \node[cell,draw=none,above right=of diag] (diag2) {\rotatebox{90}{$\ddots$}};
      \node[cell,draw=none,right=of topleft] (x) {$\cdots$};
      \node[cell,right=of x] (x2) {
        $\exists r_x.f(2^n{-}1,2^n{-}1)$
        $\forall r_x.A_1 \dots \forall r_x.A_n$
        $\forall r_x.B_1 \dots \forall r_x.B_n$
        \vphantom{$r_x^-$}$\Finish_x$ $\Finish$
        $\forall r_x^-.\Finish_x$
        $\forall r_y^-.\Finish_x$
      };
      \node[cell,right=of start] (right) {
        $\exists r_x.f(2,0)$ $\exists r_y.f(1,1)$
        $\forall r_x.A_1 \dots \forall r_x.\overline{A_n}$
        $\forall r_x.\overline{B_1} \dots \forall r_x.\overline{B_n}$
        $\forall r_y.A_1 \dots \forall r_x.\overline{A_n}$
        $\forall r_y.B_1 \dots \forall r_x.\overline{B_n}$
        \vphantom{$r_x^-$}$\Finish_x$ $\Finish_y$ $\Finish$\\
        $\forall r_x^-.\Finish_x$
      };
      \node[cell,draw=none,right=of right] (right2) {$\cdots$};
      \node[cell,right=of right2] (bottomright) {
        $\exists r_y.f(2^n{-}1,1)$
        $\forall r_y.A_1 \dots \forall r_y.A_n$
        $\forall r_y.B_1 \dots \forall r_y.\overline{B_n}$
        \vphantom{$r_x^-$}$\Finish_y$ $\Finish$
        $\forall r_x^-.\Finish_x$
      };
      \node[cell,draw=none,above=of bottomright] (y) {$\vdots$};
      \node[cell,above=of y] (y2) {
        $\exists r_y.f(2^n{-}1,2^n{-}1)$
        $\forall r_y.A_1 \dots \forall r_y.A_n$
        $\forall r_y.B_1 \dots \forall r_y.B_n$
        \vphantom{$r_x^-$}$\Finish_y$ $\Finish$
        $\forall r_x^-.\Finish_x$
        $\forall r_y^-.\Finish_x$
      };
      \node[cell,above=of y2] (topright) {
        $\Finish$ \\
        $\forall r_x^-.\Finish_x$ $\forall r_y^-.\Finish_y$
      };
      
      \begin{scope}[font=\small,gray]
        \node[below=5em of start] (x0) {$0$};
        \node[right=of x0] (x1) {$1$};
        \node[right=of x1] (q) {};
        \node[right=of q] (xn) {$2^n{-}1$};
        \node[left=5em of start] (y0) {$0$};
        \node[above=of y0] (y1) {$1$};
        \node[above=of y1] (p) {};
        \node[above=of p] (yn) {$2^n{-}1$};
      
        \draw[->,every edge/.append style={edge label={$r_x$}}]
          (start) edge (right)
          (above) edge (diag)
          (topleft) edge (x)
          (x) edge (x2)
          (x2) edge (topright)
          (right) edge (right2)
          (right2) edge (bottomright)
        ;

        \draw[->,every edge/.append style={edge label={$r_y$}}]
          (start) edge (above)
          (right) edge (diag)
          (bottomright) edge (y)
          (y) edge (y2)
          (y2) edge (topright)
          (above) edge (above2)
          (above2) edge (topleft)
        ;
      \end{scope}
    \end{tikzpicture}
    \caption{The sentences encoding a solution of the tiling problem.}
    \label{fig:tiling}
  \end{figure}
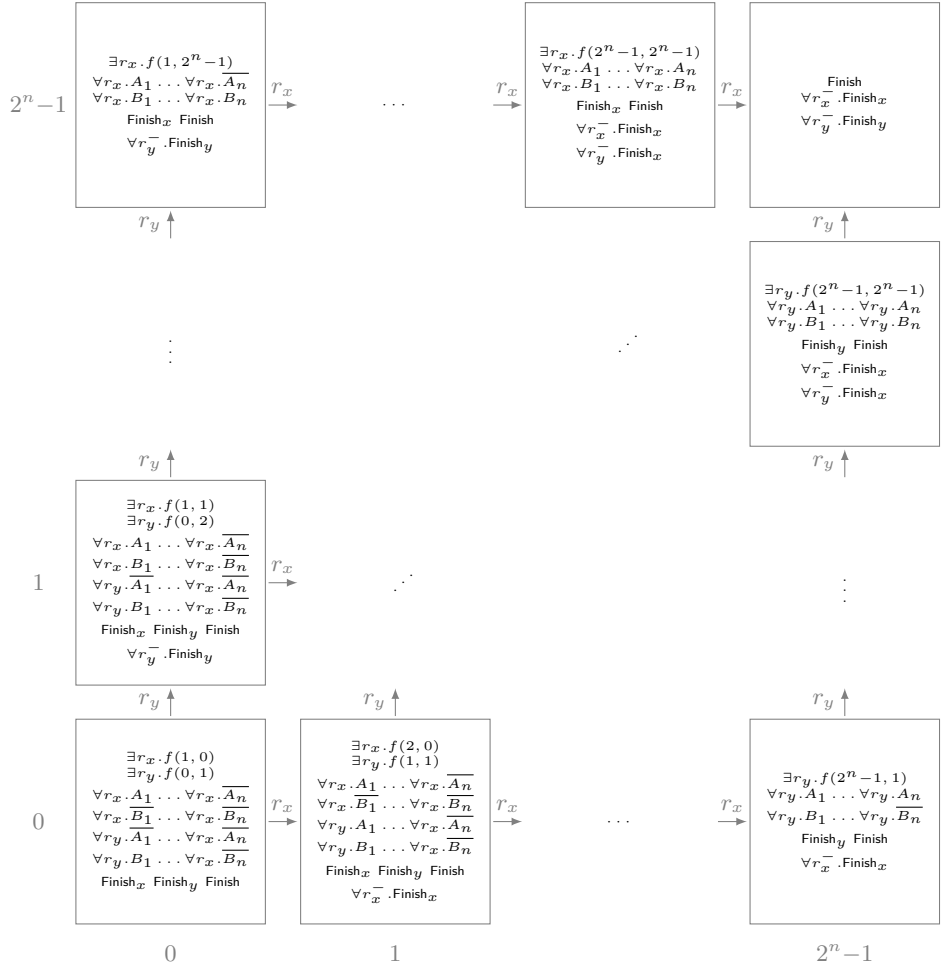
  The way the tiling is represented in the proof is sketched in Figure~\ref{fig:tiling}, which we will explain step-by-step. An arrow labeled with~$r_x$ from $(i,j)$ to $(i+1,j)$ represents a sentence of the form
  \[ C(i,j) \sqsubseteq \exists r_x. C(i+1,j), \]
  where $i\in[0,2^n-2]$ and $j\in[0,2^n-1]$, unless $i=j=0$, in which case it represents a sentence of the form 
  $\Start\sqsubseteq\exists r_x.C(1,0)$.
  We abbreviate these sentences by $H(i,j)$, and do likewise for the sentences $V(i,j)$ connecting cells $(i,j)$ and $(i,j+1)$ using $r_y$ in the vertical direction:
  \[ C(i,j) \sqsubseteq \exists r_y. C(i,j+1), \]
   where $i\in[0,2^n-1]$ and $j\in[0,2^n-2]$, unless $i=j=0$ in which case it represents $\Start\sqsubseteq\exists r_y.C(0,1)$.

  The concepts $X$ on the cells $(i,j)$ shown in Figure~\ref{fig:tiling} are those for which we will derive sentences $C(i,j) \sqsubseteq X$. For clarity, this figure does not include the sentences where $X$ is one of the conjuncts of $C(i,j)$.
  
  We construct the proof in four phases, using $m_1$, $m_2$, $m_3$ and $m_4$ sentences, respectively. Phases~(I)--(III) encode the tiling solution based on $\Tmc_1$ in a proof for $\Start\sqsubseteq\Finish$, Phase~(IV) takes care of $\Tmc_2$, completing the proof to a proof for $\Start\sqsubseteq\Success$.
  In Phase~(I), we add all sentences of~$\Tmc_1$  to the proof, except that for~\eqref{a:last-tile}, we choose only the sentence with $t=f(2^n-1,2^n-1)$, \ie the tile on the upper right cell of the grid (see Figure~\ref{fig:tiling}).
  This amounts to exactly $m_1=4n^2+6n+|h|+|v|+7$ sentences.
  
  
  In Phase~(II), we derive all necessary sentences $H(i,j)$ and $V(i,j)$ for the whole grid.
  For $H(i,j)$, we first consider the special case where $i=j=0$. In this case, we have the sentences $\Start\sqsubseteq X$ for every $X$ occurring in $C(0,0)$. Using one of the sentences \eqref{a:keep-y}--\eqref{a:end-count-y} introduced in Phase~(I), we can then use $\mathsf{CR2}$ to obtain exactly $2n$ sentences of the form
  \[ \Start\sqsubseteq\forall r_x.Y, \]
  where $Y$ ranges over the conjuncts in $[A=1]\sqcap[B=0]$, and one sentence
  \[ \Start\sqsubseteq\exists r_x.f(1,0) \]
  via~\eqref{a:x-tiling} since $\tup{f(0,0),f(1,0)}\in h$. Finally, we use a sequence of applications of~$\mathsf{CR4}$ to obtain $H(0,0)$ through $2n$ intermediate sentences that are also unique to this part of the proof. Thus, $H(0,0)$ is proven using $4n+1$ additional sentences.
  
  For the remaining $H(i,j)$, we proceed similarly. We first derive exactly $2n+1$ sentences of the form
  \[ C(i,j)\sqsubseteq X, \]
  where $X$ ranges over the concept names in $C(i,j)$, using $\mathsf{CR1}$. 
  These sentences may be shared with the proof of $V(i,j)$ in case that $j\le 2^n-2$. We note that $C(1,0)$ syntactically occurs in $H(0,0)$, which makes this application of $\mathsf{CR1}$ possible for $C(1,0)$. The remaining rule applications in this phase make sure that the remaining conceps $C(i,j)$ are all introduced, one after the other, so that $\mathsf{CR1}$ is applicable for all $C(i,j)$.
  Using one of the sentences \eqref{a:keep-y}--\eqref{a:end-count-y} introduced in Phase~(I), we can then use $\mathsf{CR2}$ to obtain exactly $2n$ sentences of the form
  \[ C(i,j)\sqsubseteq\forall r_x.Y, \]
  where $Y$ ranges over the conjuncts in $[A=i+1]\sqcap[B=j]$, and one sentence
  \[ C(i,j)\sqsubseteq\exists r_x.f(i+1,j) \]
  via~\eqref{a:x-tiling} since $\tup{f(i,j),f(i+1,j)}\in h$.
  These sentences are unique to $H(i,j)$.
  Using a sequence of applications of~$\mathsf{CR4}$ we then obtain $H(i,j)$ through $2n$ intermediate sentences that are also unique to this part of the proof. Thus, for each $H(i,j)$, where we do not have $i=j=0$, we use $6n+2$ sentences. 
  The sentences $V(i,j)$ are obtained in a similar way, but using $2n+1$ sentences already used for $H(i,j)$. 
  In total, Phase~(II) uses $$m_2=(2^n-1)^2(10n+3)+2(2^n-1)(6n+2)-(2n+1)$$ sentences,
  namely $(2n+1)+2(4n+1)=(10n+3)$ for each inner cell ($i,j\in[0,2^n-2]$), where the sentences derived using $\mathsf{CR1}$ are shared between $H(i,j)$ and $V(i,j)$,
  and $$(2n+1)+(2n+1)+2n=(6n+2)$$ for each \enquote{border} cell with either $i=2^n-1$ or $j=2^n-1$ (but not both), where only one of $H(i,j)$ or $V(i,j)$ needs to be derived. To get $H(0,0)$ and $V(0,0)$, we need $8n+2$ sentences in complete, which are $2n+1$ less than for the other cells, which is why this number is subtracted in the end.
  
  In Phase~(III), we derive all sentences of the form
  \[ C(i,j)\sqsubseteq\Finish. \]
  We start with~\eqref{a:last-tile} and~\eqref{a:back-propagate-x}, which can be combined into
  \[ C(2^n-1,2^n-1)\sqsubseteq\forall r_x^-.\Finish_x \]
  using $\mathsf{CR2}$, which together with $H(2^n-2,2^n-1)$ yields
  \[ C(2^n-2,2^n-1)\sqsubseteq\Finish_x \]
  by $\mathsf{CR3}$, and finally
  \[ C(2^n-2,2^n-1)\sqsubseteq\Finish \]
  by $\mathsf{CR2}$ and~\eqref{a:last-row} together with the sentences $C(2^n-2,2^n-1)\sqsubseteq B_i$, $1\le i\le n$, from Phase~(I).
  We can use this sentence in analogy to~\eqref{a:last-tile} to continue this process until we reach $C(0,2^n-1)\sqsubseteq\Finish$.
  Thus, we have labeled all cells in the last row of Figure~\ref{fig:tiling} with \Finish, and we can proceed similarly with the last column.
  
  For the remaining cells we can proceed similarly, but need to derive both $\Finish_x$ and $\Finish_y$ each time.
  Let $0\le i,j\le 2^n-2$, $i\neq 0$ or $j\neq 0$, and assume that we have already derived $C(i+1,j)\sqsubseteq\Finish$ and $C(i,j+1)\sqsubseteq\Finish$ for the right and upper neighbor of cell~$(i,j)$, respectively.
  We next derive
  \[
    C(i+1,j) \sqsubseteq \forall r_x^-.\Finish_x
    \quad \text{and} \quad
    C(i,j+1) \sqsubseteq \forall r_y^-.\Finish_y
  \]
  using $\mathsf{CR2}$ and Sentences~\eqref{a:back-propagate-x} and~\eqref{a:back-propagate-y}, respectively.
  We then obtain
  \[
    C(i,j) \sqsubseteq \Finish_x
    \quad \text{and} \quad
    C(i,j) \sqsubseteq \Finish_y
  \]
  via $\mathsf{CR3}$ using $H(i,j)$ and $V(i,j)$, respectively.
  Finally, $\mathsf{CR2}$ allows us to derive
  \[ C(i,j) \sqsubseteq \Finish \]
  together with~\eqref{a:combine-x-y}.
  
  For the case of $i=j=0$, similar applications result in the desired sentence $\Start\sqsubseteq\Finish$.
  %
  The total number of new sentences Phase~(III) is $m_3=6(2^n-1)+5(2^n-1)^2$, namely $3$ for each cell in the last row or the last column of the grid and $5$ for each inner cell.
  
  \begin{figure}
   \input{figures/nexptime-reduction-tbox2}
   \caption{TBox $\Tmc_2$ enforcing the use of sentences~\eqref{a:x-tiling} and~\eqref{a:y-tiling}.}
   \label{fig:nexptime-reduction-tbox2}
  \end{figure}
  
  In Phase~(IV), we now prove $\Start\sqsubseteq\Success$, making sure that really every sentence introduced in Phase~(I) is required for the proof. For this, we need to prove $\Start\sqsubseteq H_i$ for every horizontal tiling condition $h_i$, and $\Start\sqsubseteq V_i$ for every vertical tiling condition $v_i$. We infer $\Start\sqsubseteq H_i$ by first inferring $t\sqsubseteq\exists r_x.(t'\sqcap H_i)$ for $h_i=\tup{t,t'}$ using~\eqref{eq:first-h} and~\eqref{a:x-tiling}, then $t\sqsubseteq H_i$ using~\eqref{eq:second-h}, $t\sqsubseteq\forall r^-.H_i$ using~\eqref{eq:third-h}, and finally $\Start\sqsubseteq H_i$ using~\eqref{eq:every-tile}. For $\Start\sqsubseteq V_i$, we proceed similarly. In complete, for both $\Start\sqsubseteq H_i$ and $\Start\sqsubseteq V_i$, $\lvert T_I \rvert + 7\cdot(\lvert v\rvert+\lvert h\rvert)$ sentences are needed (the sentences~\eqref{eq:every-tile} only need to be counted once).%
 Note that also every sentence of the form~\eqref{a:x-tiling} and~\eqref{a:y-tiling} must be used in this proof. The final proof for $\Start\sqsubseteq\Success$ needs 2 additional sentences (\eqref{eq:new-finish} and the conclusion), so that Phase~(IV) needs $m_4=\lvert T_I\rvert + 7\cdot (\lvert v\rvert + \lvert h\rvert) + 2$ additional sentences in complete. We obtain that the complete proof for $\Start\sqsubseteq\Success$ has a size of $m$.

  $(\Rightarrow)$:
  Assume that there is a proof of $\Tmc\models\Start\sqsubseteq\Finish$ that uses $m=m_1+m_2+m_3+m_4$ sentences or less.
  Any such proof must follow the general scheme laid out above, since one cannot get from $C(2^n-1,2^n-1)\sqsubseteq\Finish$ to $C(0,0)\sqsubseteq\Finish$ without going through all tiles $(i,j)\in[0,2^n-1]\times[0,2^n-1]$.
  Hence, one needs to derive appropriate sentences $H(i,j)$ and $V(i,j)$ first, for some choice of tile types.
  It can also be verified that none of the individual derivation steps can be achieved using fewer intermediate sentences. 
  Spcifically, every sentence~\eqref{a:x-tiling} and~\eqref{a:y-tiling} must be used exactly once in such a proof, due to Phase~(IV).
  Thus, the tight bound of~$m$ requires that, for each cell~$(i,j)$, at most one concept of the form $C(i,j)$ occurs in the proof, \ie each cell is assigned a unique tile $f(i,j)\in T$.
  This yields a solution of the tiling problem due to the horizontal and vertical compatibility relations encoded in~\eqref{a:x-tiling} and~\eqref{a:y-tiling}.
  \end{proof}

Now we can proceed to the tree-shaped proof representation (see Table~\ref{tab:results2}).

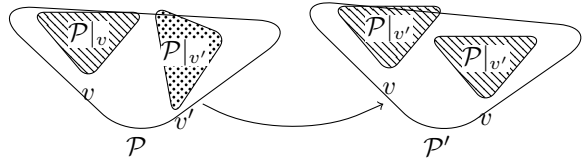
\begin{figure}[tb]
\centering
\begin{tikzpicture}[scale=0.8,block/.style={regular polygon, regular polygon sides=3,
              draw, fill=white, text width=0.5em,
              inner sep=1mm, outer sep=0mm,
              shape border rotate=-60,rounded corners=5pt}]
\path (-0.3,3) node[block, yscale=0.82, xscale=1.2, pattern=north west lines,text width=0.9em,label=below:$v$] (x1) {}
		node[inner sep=0pt,fill=white] at (x1) {$\p|_v$}
		(1.3,2.7) node[block, yscale=1.2, xscale=0.8, shape border rotate=-75, pattern=crosshatch dots,text width=0.9em,label=below:$v'$] (x2) {}
		node[inner sep=0pt,fill=white] at (x2) {$\p|_{v'}$};
\draw[rounded corners=18pt] ($(x1)+(-1.7,0.5)$) -- ($(x1)!0.5!(x2)+(0,-1.7)$) node(c1) {$\p$} -- ($(x2)+(2,0.5)$)--cycle;		
\path (4.7,3.1) node[block, yscale=0.82, xscale=1.2, pattern=north west lines,text width=0.9em,label=below:$v$] (x1') {}
	    node[inner sep=0pt,fill=white] at (x1') {$\p|_{v'}$}
		(6.3,2.6) node[block, yscale=0.82, xscale=1.2, pattern=north west lines,text width=0.9em,label=below:$v$] (x2') {}
		node[inner sep=0pt,fill=white] at (x2') {$\p|_{v'}$};
\draw[rounded corners=18pt] ($(x1')+(-1.7,0.5)$) --  ($(x1')!0.5!(x2')+(0,-1.7)$) node(c2) {$\p'$} -- ($(x2')+(2,0.5)$)--cycle;	
\draw[rounded corners=30pt, ->] ($(c1)+(1.1,0.7)$) --  ($(c1)!0.5!(c2)$) -- ($(c2)+(-0.9,0.7)$);		
\end{tikzpicture}
\caption{Transformation made in the proof of Lemma~\ref{lem:regular} for $v,v'\in V$ with $\el(v)=\el(v')$}\label{fig:lem:regular}
\end{figure}

\lemTreeProofsHyperpaths*

We first prove an auxiliary result that allows us to restrict the search to tree proofs of a specific form, in a way similar to Lemma~\ref{lem:adm-hyperproof-inside}. Given a tree proof $\p=(V,E,\el)$ and $v\in V$, we denote by $\p|_v$ the subtree of~\p with root~$v$ (which is a proof of~$\el(v)$). Moreover, \p is called \emph{regular} if, whenever $\el(v)=\el(v')$, then $\p|_v$ and $\p|_{v'}$ are isomorphic.
\begin{lemma}
\label{lem:regular}
  There exists a minimal tree proof for~$\Tmc\models\eta$ in~\R that is regular.
\end{lemma}
\begin{proof}
  Let $\p=(V,E,\el)$ be a minimal tree proof that is not regular.
  Then there are $v,v'\in V$ such that $\p|_v$ and $\p|_{v'}$ are not isomorphic.
  We choose $v$ such that $\ct{\p|_v}$ is minimal among all possible choices of $v$ and $v'$, and therefore $\ct{\p|_v}\le\ct{\p|_{v'}}$.
  We now follow a similar strategy as in the proof of Lemma~\ref{lem:adm-hyperproof-inside}, but instead of merging~$v$ and $v'$ we replace $\p|_{v'}$ in~\p by a copy of~$\p|_v$.
  By similar arguments as for Lemma~\ref{lem:adm-hyperproof-inside}, this results in another proof~$\p'$ that is admissible \wrt $\R(\Tmc,\eta)$, which is clearly also still a tree.
  Since \p is minimal, it is impossible that $\ct{\p|_v}<\ct{\p|_{v'}}$, which means that $\ct{\p'}=\ct{\p}$.
  
  We can continue in this fashion (depicted in Figure~\ref{fig:lem:regular}) until we obtain a regular proof with the same number of vertices as~\p.
  This process terminates since by our choice of~$v$ the subtree $\p_v$ cannot contain another vertex that violates the regularity condition, and thus by replacing $\p|_{v'}$ with $\p|_v$ we always decrease the number of pairs $(v,v')$ that violate the regularity condition.
  %
\end{proof}
Hence, in the following we can focus on regular tree proofs.
Next, we draw a connection between \emph{non-redundant} proofs and hyperpaths.

\begin{restatable}{lemma}{lemProofsHyperpaths}\label{lem:proofs-hyperpaths}
A derivation structure $\p=(V,E,\el)$ over \Tmc is a non-redundant proof for~$\Tmc\models\eta$ iff $\p^*$ is a hyperpath from $s$ to a vertex $v_\eta$ \st $\el(v_\eta)=\eta$.
\end{restatable}
\begin{proof}
In order to show that $\p^*$ is a hyperpath, we note that, trivially, $s,v_\eta \in V^*$. Now, for any $v\in V$, we show that $v$ is reachable from $s$ in $\p^*$. All leaves of~$\p$ and those vertices~$v$ with an \enquote{empty} incoming edge $(\emptyset,v)$ are reachable from $s$ in~$\p^*$ by the edges we added to~$E^*$. Moreover, all other vertices are reachable from at least one such leaf.
There can be no cycle through~$s$, because it has only outgoing edges. Hence, any cycle in~$\p^*$ would contradict the acyclicity of~$\p$.
%
Finally, minimality of~$\p^*$ follows from non-redundancy of~\p.

For the opposite direction, if the hyperpath $\p^*$ from $s$ to $v_\eta$ contains another sink, we can build a smaller hypergraph $\p'$ by exhaustively removing sinks $v\neq v_\eta$ and all their incoming edges. This hypergraph is also a hyperpath from $s$ to $v_\eta$, which contradicts the minimality condition in Definition~\ref{def:hyperpath} for $\p^*$. Acyclicity of the proof $\p$ follows from the acyclicity of $\p^*$.
Non-redundancy follows from the minimality of hyperpaths.
\end{proof}

We can now finish the proof of Lemma~\ref{lem:tp-hyp} by noting that, for any hyperpath~$\p^*$ as in Lemma~\ref{lem:proofs-hyperpaths}, we can unravel the corresponding proof~\p into a regular tree proof~$\p'$ (thereby recursively introducing copies of subtrees that are used in multiple inference steps). A homomorphism $h\colon\p'\to\ds$ can be defined accordingly. Since the recursive definition of $W_{\p^*}(\p^*)$ adds a summand of~$1$ for every time a vertex occurs in an edge in~$\p^*$, we have $\ct{\p'}=W_{\p^*}(\p^*)$.
Conversely, for a regular tree proof~$\p'$ with a homomorphism $h\colon\p\to\ds$, we can find a corresponding non-redundant proof~$\p$ in~\ds and obtain a hyperpath~$\p^*$ with $W_{\p^*}(\p^*)=\ct{\p'}$ by the same arguments as above.

We can show a matching lower bound by reducing a standard reasoning problem to $\MTP(\R)$.

\thMTPPolyPHard*
\begin{proof}
  We reduce the entailment problem of a GCI $A\sqsubseteq B$ from an \EL-TBox~\Tmc in, which is \PTime-hard already when $A$ and $B$ are concept names and \Tmc does not contain existential restrictions~$\exists r.D$, because then \EL is equivalent to propositional Horn logic~(see~\cite{DBLP:journals/jlp/DowlingG84} for details).
  We will reduce this problem to $\MTP(\text{ELK})$, where $n$ is encoded in unary and we can restrict our attention to the rules from Figure~\ref{fig:cr} that do not use existential restrictions.
  The challenge is to find~$\Tmc'$ and~$n$ such that $\Tmc'\models A\sqsubseteq B$ holds and there is a tree proof~\p for~$A\sqsubseteq B$ over~$\Tmc'$ with $\ct{\p}\le n$ iff $\Tmc\models A\sqsubseteq B$.
  According to~\cite{DBLP:journals/jar/KazakovKS14}, the size of $\text{ELK}(\Tmc,A\sqsubseteq B)$ is bounded polynomially in~$|\Tmc|$.
  However, for our purposes, we need to show that the minimal tree proof admissible \wrt this derivation structure is similarly bounded.
  
  \begin{claim}
    There is a polynomial~$p$ such that, for any TBox~$\Tmc$ and GCI $A\sqsubseteq B$ as above, the minimal tree proof for $A\sqsubseteq B$ over~\Tmc that is admissible \wrt $\text{ELK}(\Tmc,A\sqsubseteq B)$ satisfies $\ct{\p}\le p(|\Tmc|)$.
  \end{claim}
  \noindent
  \emph{Proof of claim.}
  Since \Tmc does not contain existential restrictions, we can assume that all GCIs in~\Tmc are of the form $A_1\sqcap\dots\sqcap A_m\sqsubseteq B_1\sqcap\dots\sqcap B_k$ (*), where $A_1,\dots,A_m,B_1,\dots,B_k$ are concept names, the order and multiplicity of concept names is irrelevant, and we view $\top$ as the \enquote{empty} conjunction.
  The basic idea is to use \emph{unit propagation}~\cite{DBLP:journals/jlp/DowlingG84} as for Horn clauses to derive all consequences, including $A\sqsubseteq B$.
  That is, starting from GCIs of the form $A\sqsubseteq B_1\sqcap\dots B_k$, we \enquote{propagate} the GCIs $A\sqsubseteq B_j$ to other GCIs where $B_j$ occurs on the left-hand side.
  Once we have derived GCIs $A\sqsubseteq A_i$ for all concept names~$A_i$ on the left hand side of a GCI $A_1\sqcap\dots\sqcap A_m\sqsubseteq C_1\sqcap\dots C_k$, then we can also \enquote{propagate} the GCIs $A\sqsubseteq C_j$, and so on until we reach $A\sqsubseteq B$.
  However, we have to choose a strategy that results in a polynomially sized proof tree, which means that we need to avoid using a derived GCI in several following inference steps to avoid duplicating the whole subproof in the tree.
  
  We use the following strategy to compute a tree proof for~$\Tmc\models A\sqsubseteq B$:
  \begin{enumerate}
    \item Initialize the conjunction~$X:=A$ of already \enquote{propagated} concept names, for which we can derive the GCI $A\sqsubseteq X$ using the inference rule~$\mathsf{R}_0$.
    \item Select all GCIs of the form (*) from~\Tmc for which all of $A_1,\dots,A_m$ occur in~$X$.
    \item Extend these GCIs to $X\sqsubseteq B_1\sqcap\dots\sqcap B_k$ by first deriving $X\sqsubseteq X$ via~$\mathsf{R}_0$, from this $X\sqsubseteq A_1\sqcap\dots\sqcap A_m$ by splitting $X$ on the right-hand side using~$\mathsf{R}_\sqcap^-$, and then combining this GCI with the original GCI of the form~(*) by~$\mathsf{R}_\sqsubseteq$.
    \item Combine all selected GCIs, which now have the same left-hand side~$X$, into a single sentence $X\sqsubseteq Y$, where $Y$ is the conjunction of all concept names~$B_j$ on the right-hand sides of the selected GCIs, using the rule~$\mathsf{R}_\sqcap^+$.
    \item Combine this GCI with $X\sqsubseteq X$ (from~$\mathsf{R}_0$) to obtain $X\sqsubseteq X\sqcap Y$ via~$\mathsf{R}_\sqcap^+$.
    \item Derive $A\sqsubseteq X\sqcap Y$ from $A\sqsubseteq X$ and $X\sqsubseteq X \sqcap Y$ by~$\mathsf{R}_\sqsubseteq$.
    \item Continue with Step~2 $X\sqcap Y$ in place of~$X$.
  \end{enumerate}
  It can be checked that the resulting tree proof derives most sentences only once, except for ones that can be derived using a linear proof (where each hyperedge $(S,d)$ satisfies $|S|=1$) of polynomial size.
  For example, the GCI $X\sqsubseteq A_1\sqcap\dots\sqcap A_m$ may be derived several times if several GCIs in~\Tmc have the same left-hand side, but this cannot cause an exponential growth of this tree \wrt the size of~\Tmc.
  
  \medskip
  Taking the polynomial~$p$ from our claim, we set $n:=p(|\Tmc|)$ and
  \[ \Tmc' := \Tmc \cup \{ A\sqsubseteq A_1,\ A_1\sqsubseteq A_2, \dots, A_n\sqsubseteq B\}, \]
  where $A_1,\dots,A_n$ are concept names not occurring in~\Tmc.
  Clearly, $\Tmc'\models A\sqsubseteq B$ and the existence of a tree proof~\p for $A\sqsubseteq B$ over~$\Tmc'$ with $\ct{\p}\le n$ is equivalent to $\Tmc\models A\sqsubseteq B$, since any proof that uses the new concept names must use at least $n+1$ sentences.
  Moreover, we can compute $n$ (in binary representation) and output it in unary representation using a logarithmically space-bounded Turing machine, and similarly for~$\Tmc'$.
  Hence, the above construction constitutes the desired \LogSpace-reduction.
\end{proof}

We conjecture that an \ExpTime-lower bound for $\MTP^\exp_\bin(\R)$ could be shown in a similar fashion, by using a binary counter to enforce an artificial proof of $A\sqsubseteq B$ that has exponentially many vertices. However, so far we were not successful in finding a concrete exponential \reasoner{} where we could show an exponential upper bound on the size of minimal tree proofs.

We show \NP-hardness for finding a tree proof (of size less than $n$) for \ExpTime-complete \ELI, the extension of \EL by inverse roles.
%

\thMTPExpNPHard*
\begin{proof}
We show hardness using the \reasoner that uses the \ELI-calculus shown in Figure~\ref{fig:ELIcr}, thus showing that the problem is already hard for that specific \reasoner.
We will provide a translation of the SAT problem to $\MTP^\exp_\un(\R)$. In SAT, we are given a Boolean formula of the form $\varphi=c_1\wedge \dots \land c_m$ over a finite set of variables $\{x_1,\dots, x_n\}$, where each $c_i$ is a disjunction of 
literals (variables or its negations), and we want to decide whether $\varphi$ is decidable.

Without loss of generality, we assume that for every variable $x_i$, $1\leq i\leq n$, we have a corresponding clause $x_i\vee\neg x_i$. Given such a formula $\varphi$, we construct a TBox $\Tmc_\varphi$ s.t. $\Tmc_\varphi\models A\sqsubseteq F$ can be shown with a tree proof of size $2n+4m+9$ iff $\varphi$ is satisfiable. 

For every variable $x_i$, $1\leq i\leq n$, we use the following sentences which are supposed to ``guess'' the value of $x_i$:
\begin{align*}
 A&\sqsubseteq \forall r.X_i,  &&A\sqsubseteq \forall r.\overline{X}_i.
\end{align*}
For every clause $c_j$, $1\leq j\leq m$, we add 
\begin{align*}
X_i&\sqsubseteq\forall r.C_j, &&\text{ for every literal of the form } x_i\in c_j, \text{ and} \\
\overline{X}_i&\sqsubseteq\forall r.C_j, &&\text{ for every literal of the form }\neg x_i\in c_j.
\end{align*}
In addition, we use the following sentences whose meaning will become clear below.
\begin{align*}
 A &\sqsubseteq\exists r.B,  && B\sqsubseteq\exists r.\top, \\
 C_1\sqcap\ldots\sqcap C_m &\sqsubseteq\forall r^-.F, 
  && F\sqsubseteq\forall r^-.F
\end{align*}

We show that $\varphi$ is satisfiable iff $\Tmc_\varphi\models A\sqsubseteq F$ can be shown using a tree proof with $2n+4m+9$ vertices.

$(\Rightarrow)$: We first show that, given an assignment $v\colon\{x_1,\ldots,x_n\}\to\{0,1\}$ that makes $\varphi$ true, we can construct a tree proof for $\Tmc_\varphi\models A\sqsubseteq F$ with $2n+4m+9$ vertices. 
The assignment $v$ can be expressed as a conjunction over the set 
$$\mathcal{V}=\{X_i\mid 1\leq i\leq n,\ v(x_i)=1\}\cup\{\overline{X}_i\mid 1\leq i\leq n,\ v(x_i)=0\}.$$
In the proof, we step-wise infer the sentence $A\sqsubseteq\exists r.(B\sqcap\bigsqcap\mathcal{V})$ using rule $\mathsf{CR4}$, the sentence $A\sqsubseteq\exists r.B$, and for every $1\leq i\leq n$, the respective sentence $A\sqsubseteq\forall r.X_i$ or $A\sqsubseteq\forall r.\overline{X}_i$, depending on whether $v(x_i)=1$ or $v(x_i)=0$. This creates a subtree of the final proof of size $2n+1$. 

Since $A\sqsubseteq\exists r.(B\sqcap\bigsqcap\mathcal{V})$ can be derived, we can use $\mathsf{CR1}$ to infer $B\sqcap\bigsqcap\mathcal{V}\sqsubseteq B$. Using $\mathsf{CR2}$ on this sentence and $B\sqsubseteq\exists r.\top\in\Tmc_\varphi$, we infer $B\sqcap\bigsqcap\mathcal{V}\sqsubseteq\exists r.\top$. Starting from this sentence, we step-wise use $\mathsf{CR4}$ to infer 
\[
 B\sqcap\bigsqcap\mathcal{V}\sqsubseteq\exists r.(C_1\sqcap\ldots\sqcap C_m),
\]
where in each step, we use for a different clause $C_j$, $1\leq j\leq m$, some sentence $\overline{X}_i\sqsubseteq \forall r.C_j\in\Tmc_\varphi$ or $X_i\sqsubseteq \forall r.C_j\in\Tmc_\varphi$, depending on whether $v(x_i)=0$ or $v(x_i)=1$. This construction results in another subtree of the proof which has a size of $4m+3$ (note that for each step, we need to derive $B\sqcap\bigsqcap\mathcal{V}\sqsubseteq X_i$ (or $B\sqcap\bigsqcap\mathcal{V}\sqsubseteq\overline{X}_i$) and $B\sqcap\bigsqcap\mathcal{V}\sqsubseteq \forall r.C_j$ as intermediate axioms to be able to apply $\mathsf{CR4}$).
%
We further extend this tree by inferring $B\sqcap\bigsqcap\mathcal{V}\sqsubseteq F$ using $C_1\sqcap\ldots\sqcap C_m\sqsubseteq\forall r^-.F\in\Tmc_\varphi$ and $\mathsf{CR3}$. This uses another 2 vertices, so that the resulting sub-tree of the proof has size $4m+5$. Finally, we use $\mathsf{CR2}$ on this sentence and $F\sqsubseteq\forall r^-.F\in\Tmc_\varphi$, resulting in a sub-tree of size $4m+7$ with its root labeled $B\sqcap\bigsqcap\mathcal{V}\sqsubseteq\forall r^-.F$. 

We now have one sub-tree of size $2n+1$ with its root labelled $A\sqsubseteq\exists r.(B\sqcap\bigsqcap\mathcal{V})$, and one subtree of size $4m+7$ with its root labeled with $B\sqcap\bigsqcap\mathcal{V}\sqsubseteq\forall r^-.F$. We connect both subtrees using $\mathsf{CR3}$, resulting in a proof for $A\sqsubseteq F$ of size $2n+4m+9$. 

$(\Leftarrow)$: For the other direction, assume that $\Tmc_\varphi\models A\sqsubseteq F$ can be shown using a proof of size $2n+4m+9$. We demonstrate how we can construct from such proof an assignment $v\colon\{x_1,\ldots,x_n\}\to\{0,1\}$ that satisfies $\varphi$. A proof of $\Tmc_\varphi\models A\sqsubseteq F$ needs to infer at some point sentences of the following form:
\begin{align*}
A&\sqsubseteq\exists r.C,  &&C\sqsubseteq\forall r^-.F, \\ 
C&\sqsubseteq F, &&C\sqsubseteq\exists r.(C_1\sqcap\ldots\sqcap C_m).
\end{align*}
The sentences in $\Tmc_\varphi$ only allow to infer $A\sqsubseteq\exists r.C$ provided that $C$ is a conjunction of $B$ with concept names from the set $\{X_1$, $\overline{X}_1$, $\ldots$, $X_n$, $\overline{X}_n\}$. 
The shape of $C$ is further restricted by the fact that we can only infer $C\sqsubseteq\exists r.(C_1\sqcap\ldots\sqcap C_m)$ provided that $C$ contains $B$ as conjunct, and at least one conjunct of the form $X_i$ or $\overline{X}_i$ for each variable $x_i$, $1\leq i\leq n$. The latter is enforced since for every such variable $x_i$, there is some clause $c_j=x_i\vee \neg x_i$, so that $C_j$ occurs under a universal restriction only in the sentences $X_i\sqsubseteq\forall r.C_j$ and $\overline{X}_i\sqsubseteq\forall r.C_j$ in $\Tmc_\varphi$. 

A proof for $C\sqsubseteq\exists r.(C_1\sqcap\ldots\sqcap C_m)$ requires at least $4m+3$ vertices (1 for $C\sqsubseteq B$, 2 for $C\sqsubseteq\exists r.\top$, and then we need to derive each $C\sqsubseteq\forall r.C_j$ by $\mathsf{CR2}$ from $C\sqsubseteq X_i$ / $C\sqsubseteq \overline{X}_i$ and $X_i\sqsubseteq\forall r.C_j$ / $\overline{X}_i\sqsubseteq\forall r.C_j$, and another $m$ steps of $\mathsf{CR4}$ to obtain the goal). Inferring from this $C\sqsubseteq F$ and then $C\sqsubseteq\forall r^-.F$ requires 2 more vertices each, so that at this point, at least $4m+7$ vertices of the proof have been assigned. As we still need to infer $A\sqsubseteq F$, this leaves $2n+1$ vertices for proving $A\sqsubseteq\exists r.C$. For this, we need to use the sentence $A\sqsubseteq\exists r.B$ once, and 2 vertices for every additional conjunct in $C$, meaning that $C$ has at most $n+1$ conjuncts. $C$ is a conjunction of $B$ and at least one of $X_i$ and $\overline{X}_i$ for each $x_i$ $1\leq i\leq n$. As there are at most $n$ conjuncts, this means that for every $x_i$, $1\leq i\leq n$, $C$ contains exactly one conjunct, based on which we define the valuation $v$: $v(x_i)=0$ if $\overline{X}_i$ occurs as conjunct, and $v(x_i)=1$ if $X_i$ occurs as conjunct. We can furthermore show that $v$ satisfies each clause $c_j$, $1\leq j\leq m$: the proof used some sentence of the form $X_i\sqsubseteq\forall r.C_j$ or $\overline{X}_i\sqsubseteq\forall r.C_j$; in the first case, we have $v(x_i)=1$ by definition of $v$ ($X_i$ must be a conjunct of $C$), and $x_i\in c_j$ by construction of $\Tmc_\varphi$, and in the latter, we have $v(x_i)=0$ by the definition of $v$ and $\neg x_i\in c_j$ by the construction of $\Tmc_\varphi$. It follows that $v$ is a satisfying assignment of $\varphi$. 

We have shown that $\varphi$ is satisfiable iff $\Tmc_\varphi\models A\sqsubseteq F$ has a tree proof of size at most $2n+4m+9$, which means that SAT can be reduced to $\MTP^\exp_\un(\R)$. Consequently, $\MTP^\exp_\un(\R)$ is \NP-hard.
\end{proof}

We can show the remaining upper bound in a similar way as before.
\thMTPExpUnaryNPMem*
\begin{proof}
  We can use the same arguments as in the proof of Theorem~\ref{th:exp-unary-NP-mem}, but allow to guess vertices of $\R(\Tmc,\eta)$ several times and additionally check that we obtain a tree proof.
  %
\end{proof}

Finally, in the next theorem we argue that soundness and completeness of \emph{the forgetting-based approach} (FBA) rely on soundness and completeness of the constituent components.

\FBA*
\begin{proof}
It is not hard to see that the hypergraph $\ds = (V, E, \el)$ constructed this way is indeed a proof according to Definition~\ref{def:proof}. First, $\ds$ is a \emph{derivation structure}: it is \emph{grounded} since every vertex $v\in V$ without an incoming edge is labeled with an sentence from $\Jmc\subseteq\Tmc$; and it is \emph{sound} since for every edge $(S,d)\in E$, $\{\el(s)\mid s\in S\}$ is a justification for $\el(d)$. Second, every $(S,d)\in E$ is obtained from a pair $\Tmc_i,\Tmc_{i+1}$ from our generated sequence of TBoxes, that is, $\{\el(s)\mid s\in S\}\subseteq\Tmc_i$ and $\el(d)\in\Tmc_{i+1}$, which means that $\ds$ is an \emph{acyclic} derivation structure. It follows that $\ds$ is a proof according to Definition~\ref{def:proof}. Thus, by Lemma~\ref{lem:proof-properties}, FBA is sound. Moreover, even if the forgetting process itself is not complete, FBA always generates a proof.
\end{proof}

\newpage
\section{Examples of Generated Proofs}\label{appendix:generated}

We finally discuss a few more examples that illustrate the differences between the three proof generators compared in Section~\ref{sec:evaluation}.
The images in Figures~\ref{fig:ex4} and \ref{fig:ex1}--\ref{fig:ex5} were generated automatically from our 
datasets. Recall that hyperedges are denoted by blue boxes labelled with the type of inference rule (for \Elk) 
or the forgotten symbol(s) (for FBA).
%
%
Figures~\ref{fig:ex1} and~\ref{fig:ex3} depict two cases where the \Elk proofs are arguably better than the ones generated by FBA, while Figures~\ref{fig:ex4} and~\ref{fig:ex5} illustrate advantages of FBA.
Note that we chose quite small proofs as examples, because larger ones would not easily fit on these pages.

\begin{figure}[hb]
  \includegraphics[width=.45\linewidth]{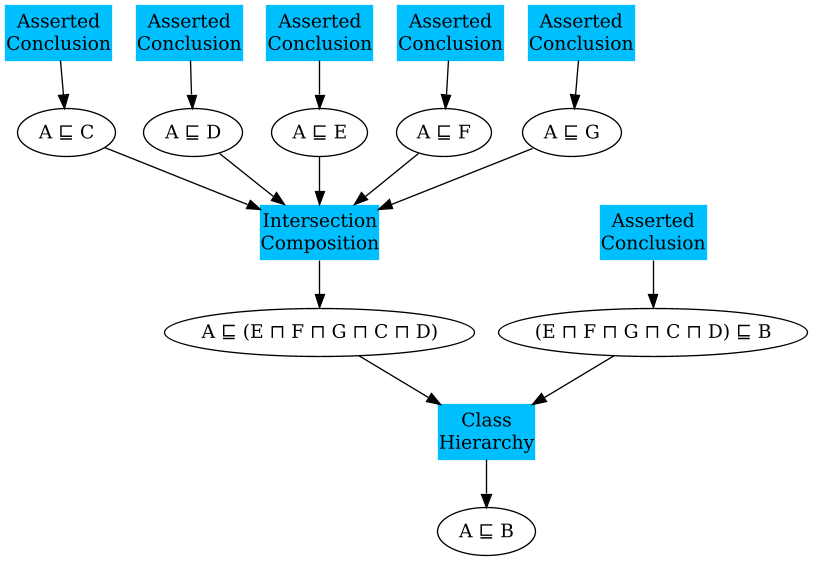}
  \hfill
  \includegraphics[width=.4\linewidth]{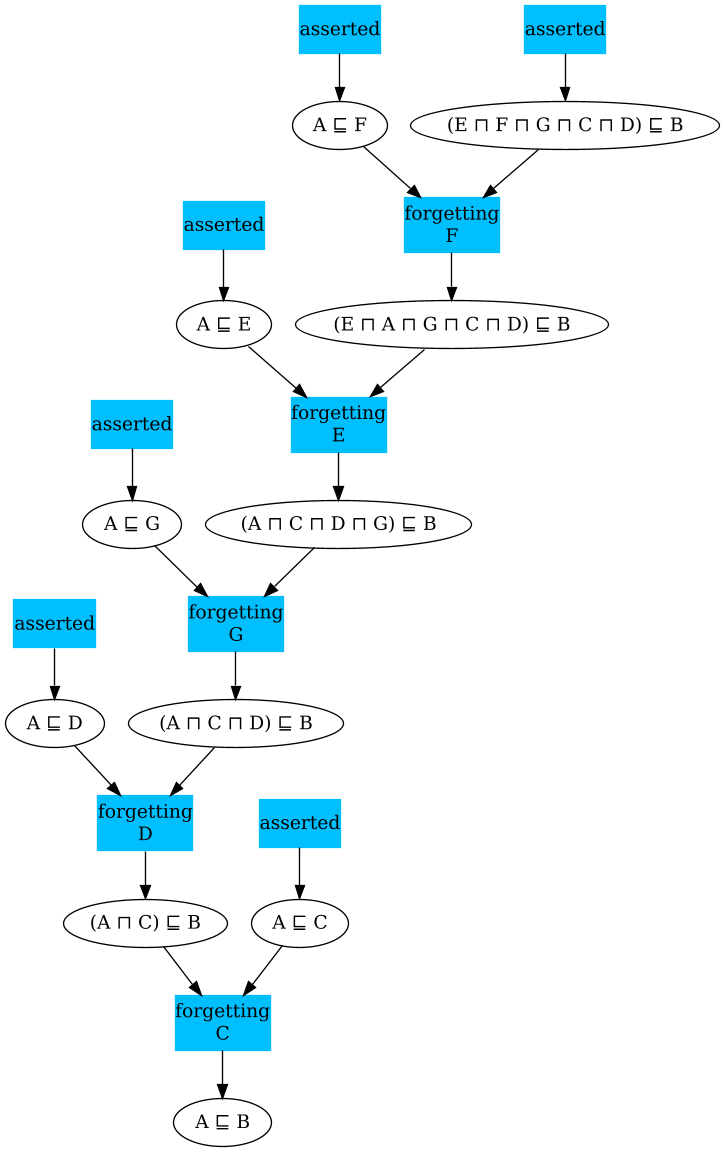}
  \caption{A proof generated by \Elk (left) vs.\ one for the same entailment generated via FBA (right; using either \Lethe or \Fame). Using the $n$-ary ``Intersection Composition'' rule, \Elk allows to formulate a cleaner proof, while the iterative forgetting approach requires separate steps to eliminate each symbol.}
  \label{fig:ex1}
\end{figure}


\begin{figure}
  \centering
  \includegraphics[width=.6\linewidth]{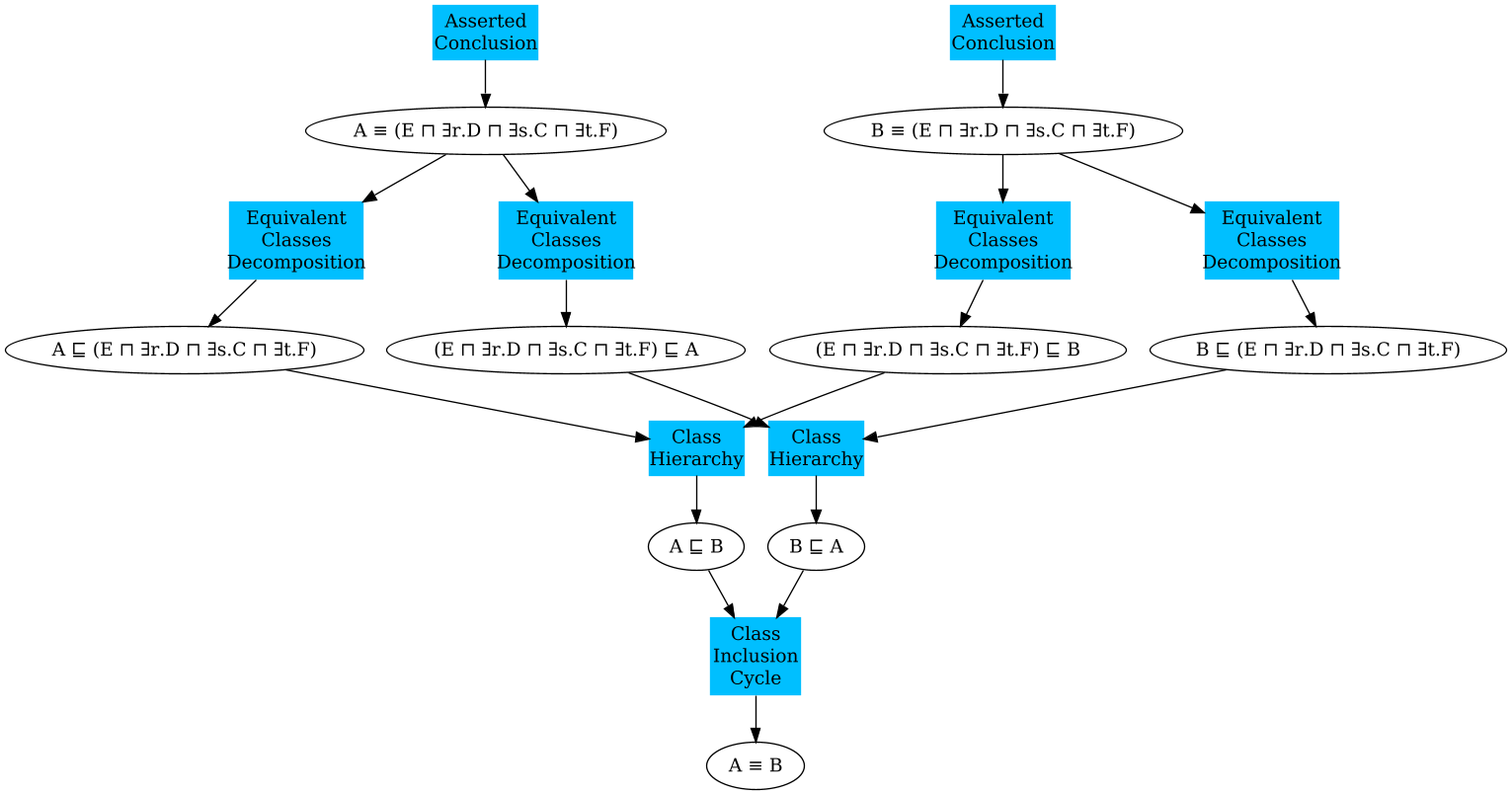}\\
  \includegraphics[width=.45\linewidth]{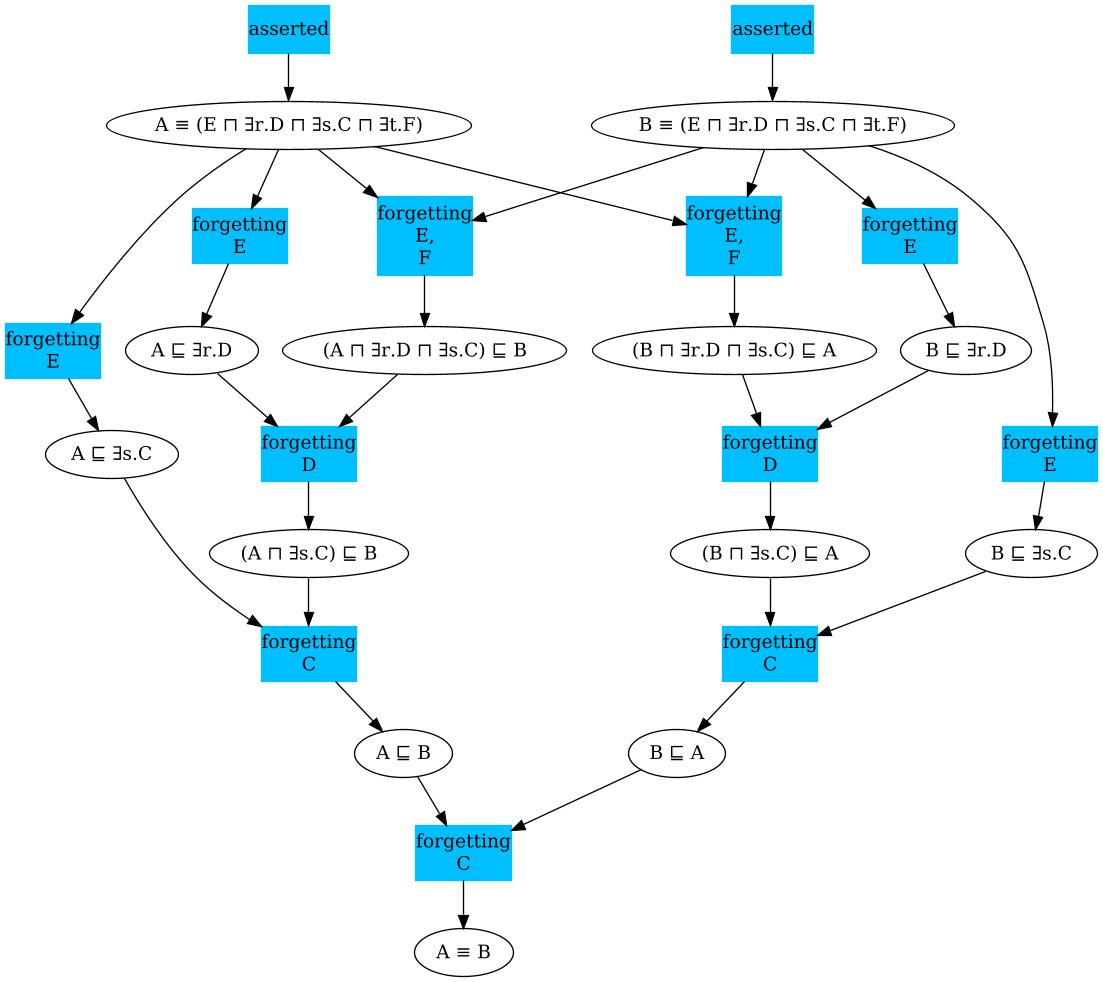}\\[1ex]
  \includegraphics[width=\linewidth]{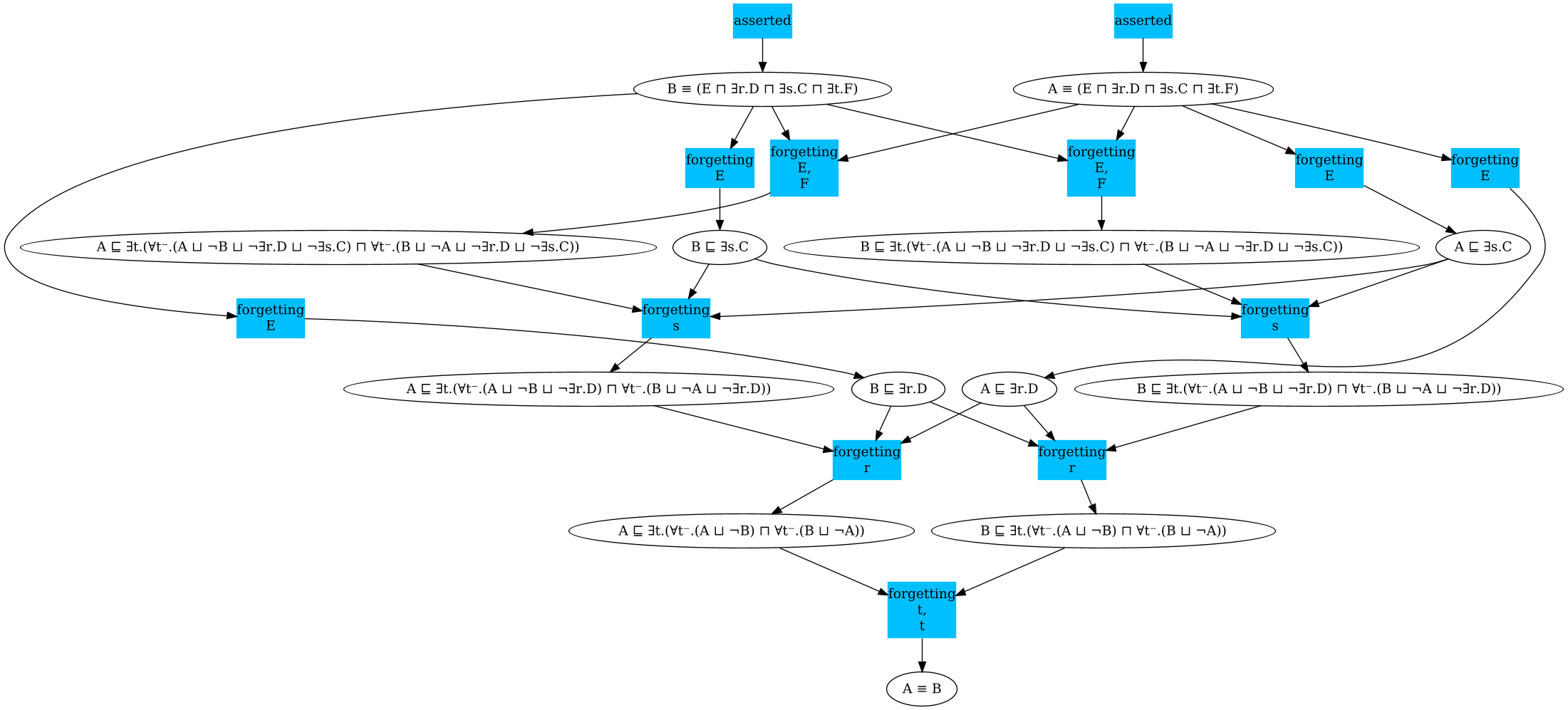}
  \caption{Again, \Elk (top) allows to eliminate several symbols at the same time, while the forgetting-based proofs (center: \Lethe, bottom: \Fame) are more complicated. \Fame makes extensive use of the expressivity of $\mathcal{ALCOI}$ because it needs to preserve all $\mathcal{ALCOI}$-consequences, which is a disadvantage in this case.}
  \label{fig:ex3}
\end{figure}


\begin{figure}
  \centering
  \includegraphics[width=.25\linewidth]{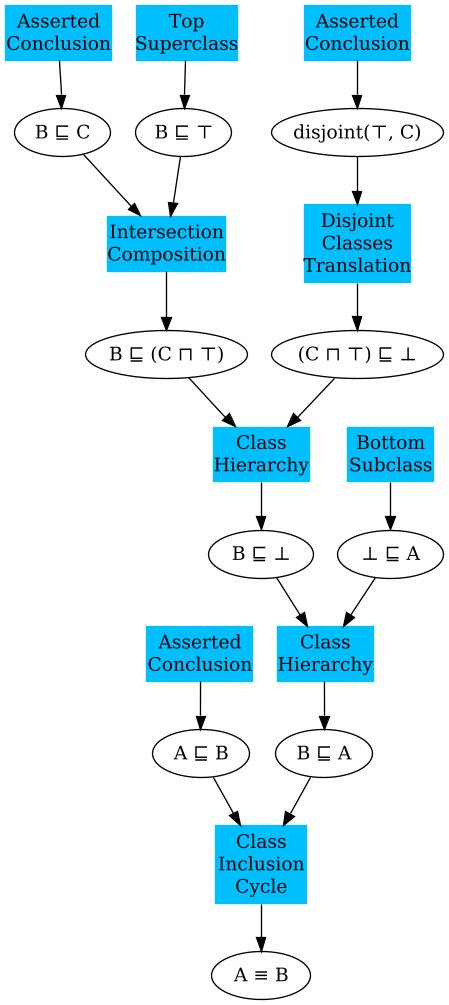}
  \qquad
  \includegraphics[width=.205\linewidth]{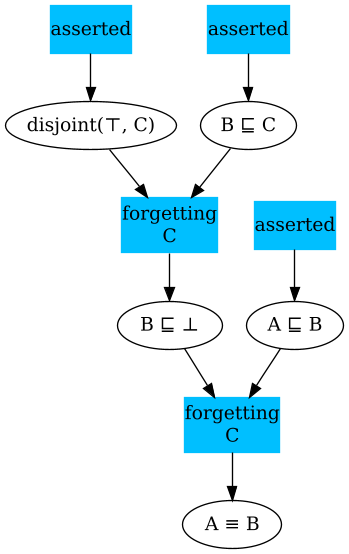}
  \caption{\Elk (left) makes many simple steps, which are collapsed by FBA (right) into a simpler proof. Assuming basic knowledge about description logics, the latter is arguably preferable.}
  \label{fig:ex5}
\end{figure}

\fi

\end{document}